\documentclass[twoside,11pt]{article}

\usepackage{blindtext}

\usepackage[preprint]{jmlr2e}

\usepackage{latexsym}
\usepackage{psfrag,epsf}
\usepackage{amsmath}
\usepackage{graphicx}
\usepackage{enumerate}
\usepackage{url} 
\usepackage[utf8]{inputenc}
\usepackage{booktabs}
\usepackage{amssymb}
\usepackage[colorinlistoftodos]{todonotes}
\usepackage{caption}
\usepackage{subcaption}
\usepackage{lscape}
\usepackage{verbatim}
\usepackage{csvsimple}
\usepackage{multirow}
\usepackage{textcomp}
\usepackage{natbib}
\usepackage{mathabx}
\usepackage{tabularx}
\usepackage[linesnumbered,ruled,vlined]{algorithm2e}
\usepackage{bm}
\usepackage{wrapfig}
\usepackage{algorithmic}
\usepackage{pdfpages,setspace}
\usepackage[normalem]{ulem}

\usepackage{enumitem}
\usepackage{mathrsfs}

\DeclareMathOperator*{\argmin}{argmin}

\newcommand{\bbL}{{\mathbb L}}

\newcommand{\ba}{{\mathbf{a}}}

\newcommand{\bdlambda}{{\boldsymbol{\lambda}}}
\newcommand{\bdbeta}{{\boldsymbol{\beta}}}

\newcommand{\bdomega}{{\boldsymbol{\omega}}}

\newcommand{\wh}{\widehat}
\newcommand{\wt}{\widetilde}
\newcommand{\E}{\mathbb{E}}
\newcommand{\pr}{\mathbb{P}}

\newcommand{\BX}{\boldsymbol{X}}

\newcommand{\bY}{\mathbf{Y}}
\newcommand{\bZ}{\mathbf{Z}}

\newcommand{\scrI}{\mathscr{I}}

\newcommand{\scrT}{\mathscr{T}}

\newcommand{\calS}{ \mathcal{S} }

\newcommand{\wsc}{ \boldsymbol{\omega}_{\calS^c} }

\newcommand{\fzeros}{ \boldsymbol{f}_{0 \calS} }

\newcommand{\Tnjj}{  \mathscr{T}_n^{(j, j)} }
\newcommand{\Tlnjj}{  \mathscr{T}_{1n}^{(j, j)} }

\newcommand{\TSS}{  \mathscr{T}^{(\calS, \calS)} }

\newcommand{\TScS}{  \mathscr{T}^{(\calS^c, \calS)} }

\newcommand{\Tjj}{  \mathscr{T}^{(j, j)} }
\newcommand{\T}{  \mathscr{T} }

\newcommand{\All}{  \mathscr{A}^{(\calS_1, \calS_1)} }
\newcommand{\Anll}{  \mathscr{A}_n^{(\calS_1, \calS_1)} }
\newcommand{\Anoo}{  \mathscr{A}_n^{(\calS_0, \calS_0)} }
\newcommand{\Anlo}{  \mathscr{A}_n^{(\calS_1, \calS_0)} }
\newcommand{\Anol}{  \mathscr{A}_n^{(\calS_0, \calS_1)} }

\newcommand{\dAnll}{  \mathscr{A}_{n, \bar{\lambda}_2}^{(\calS_1, \calS_1)} }

\newcommand{\QSS}{  \mathscr{Q}^{(\calS, \calS)} }

\newcommand{\tr}{{\rm tr}}
\newcommand{\vertiii}[1]{{\left\vert\kern-0.15ex\left\vert #1 \right\vert\kern-0.15ex\right\vert}}

\newtheorem{condition}{Condition}

\usepackage{lastpage}
\jmlrheading{23}{2022}{1-\pageref{LastPage}}{1/21; Revised 5/22}{9/22}{21-0000}{Xingche Guo and Yehua Li and Pang Du}

\ShortHeadings{MoFI-FLR: Model Form Identification in Functional Linear Regressions}{Guo and Li and Du}
\firstpageno{1}

\begin{document}

\title{Model Form Identification in High-Dimensional Functional Linear Regressions}

\author{\name Xingche Guo \email xingche.guo@uconn.edu \\
       \addr Department of Statistics\\
       University of Connecticut\\
       Storrs, CT 06269, USA
       \AND
       \name Yehua Li \email yehua.li@ucr.edu \\
       \addr Department of Statistics\\
       University of California\\
       Riverside, CA  92521, USA
       \AND
       \name Pang Du \email pangdu@vt.edu \\
       \addr Department of Statistics\\
       Virginia Tech\\
       Blacksburg, VA 24061, USA}

\editor{Rajarshi Guhaniyogi}

\maketitle

\begin{abstract}
High-dimensional functional data are becoming increasingly common in fields such as environmental monitoring and neuroimaging. This paper studies high-dimensional functional linear regression models that relate a scalar response to ultra-high-dimensional functional predictors, where each predictor is treated as a random element in an infinite-dimensional functional space. To address the dual challenges of high-dimensionality and model interpretability, we propose MoFI-FLR, a novel two-step estimation framework rooted in reproducing kernel Hilbert space (RKHS) theory. The first step employs a functional elastic-net penalty to screen out irrelevant covariates, while the second step decomposes each selected predictor's functional coefficient into an interpretable finite-dimensional simple component and an infinite-dimensional complementary complement. By penalizing only the complementary component, our method automatically distinguishes simple effects, which consist only of the simple component, from complex effects, which also include complementary deviations. Under mild regularity conditions, we establish non-asymptotic theoretical guarantees, demonstrating that MoFI-FLR consistently recovers the active covariates and accurately identifies their true functional forms. We develop a computationally efficient algorithm to implement the proposed method and evaluate its performance through comprehensive simulation studies and an application to Psychomotor Vigilance Task EEG data.

\end{abstract}


\section{Introduction}

Functional data analysis (FDA) has emerged as a vibrant research area, achieving remarkable success in a wide range of fields such as neuroscience,  environment, linguistics, medicine, and economics \citep{RamsaySilverman05, wang2016functional, LiQiuXu2022}.
Among its many developments, regression methods that relate functional predictors to scalar responses have proven especially influential and widely applied. 
There has been a large volume of literature on scalar-on-function regression models, where the most studied model is the functional linear model (FLM); see \cite{James2002, Muller2005, CaiHall06, ReissOgden2007, Crambes2009, CaiYuan2012, Lei2014, Shang-Cheng2015, Liu2021}, among others.
It is also natural to extend these ideas to settings with many functional predictors: typical of modern high‑dimensional settings where the number of predictors $p$ can exceed the sample size $n$, motivating high‑dimensional functional regression frameworks \citep{FanJamesRadchenko2015}.

We consider the following high-dimensional functional linear regression model to characterize the functional nature of
explanatory variables. 
Let $Y_i$ be the scalar response for the $i$th subject, $i=1, \ldots, n$, and $X_{ij}(\cdot)$  be the $j$th functional predictors for the $i$th subject taking values in $\mathbb{L}_2( {\cal T}_j)$, where $\mathcal{T}_j$ is a compact interval, $j = 1, \dots, p$.
The high-dimensional functional linear regression model assumes 
\begin{equation}\label{eq:aim3a}
	Y_i = \sum_{j=1}^p \int_{\mathcal{T}_j}X_{ij}(t)\beta_j(t)dt + \epsilon_i,
\end{equation}
where $\beta_j(\cdot) \in \mathbb{L}_2({\cal T}_j)$ are unknown coefficient functions, 
$\epsilon_i$ are random noises independent of
$X_{ij}(\cdot)$ for all $j$ with zero mean and bounded variance.
Without loss of generality, assume that both $Y_i$ and $X_{ij}(t)$ are centered to have mean $0$ so that no intercept is needed in \eqref{eq:aim3a}.

In the past decade, variable selection under this model or its variants has garnered considerable attention, with several examples provided below. 
\cite{reiss2010} considered a generalized functional principal component regression model where the simultaneous confidence bands for the coefficient function can be used to detect a zero or constant coefficient function. 
\cite{gertheiss2013variable} considered selecting the functional covariates in a functional linear model through a penalty combining the $L_2$-norms of coefficient functions and their second derivatives.  
\cite{matsui2011variable} and
\cite{matsui2014variable} extended the lasso and SCAD penalty, respectively, to the settings of functional linear regression models and multi-class functional logistic regression models.
\cite{collazos2016consistent} proposed a step-wise variable selection procedure for functional regression models based on likelihood ratio tests. 
\cite{kong2016partially} studied the simultaneous selection of functional covariates and scalar covariates in a high-dimensional partially functional linear regression model. 
\cite{zhang2017funnel} extended the elastic net penalty to disentangle the effects of overlapping genes in time-course gene set enrichment analysis.
\cite{XueYao2021} introduced a hypothesis‐testing framework to assess whether a given functional coefficient is identically zero.
\cite{ghosal2022variable} generalized regularized group variable selection methods such as group lasso, group SCAD, and group MCP to functional linear concurrent models. 
\cite{guo2024rkhs} propose a functional elastic-net method and establish a non-asymptotic theoretical framework for variable selection and minimax optimal prediction. The common focus of all these papers is essentially to identify zero and nonzero coefficient functions $\beta_j(t)$.

On the other hand, assessing whether a functional coefficient $\beta_j(t)$ in \eqref{eq:aim3a} follows a simple, low‑dimensional parametric form or instead requires a fully nonparametric specification is important for interpretability, parsimony, and efficient estimation. 
Below are two illustrative examples:

\noindent\textbf{Example 1 (Appliances energy use):} 
The Appliances Energy Prediction dataset analyzed by \cite{candanedo2017data} comprises recordings of house temperature, indoor humidity, and other meteorological variables from a wireless sensor network and a nearby airport station. These variables are recorded every 10 minutes so that their trajectories throughout the day can be considered as functional predictors.
Suppose the goal is to predict the total daily appliances energy consumption from these functional predictors using the high-dimensional functional linear model (\ref{eq:aim3a}).
Let $X_{i1}(t)$ denote the house temperature trajectory on day $i$, and $Y_i$ denote the total appliances energy consumption of that house on that day.
Under a parametric null hypothesis that temperature has a time‐invariant effect on energy consumption, i.e.
$
\beta_1(t)\equiv\gamma
$, for some unknown constant $\gamma$, then only the daily averaged temperature
\[
\bar X_{i1}
=\frac{1}{|\mathcal T_1|}\int_{\mathcal T_1} X_{i1}(t)\,\mathrm dt
\]
is predictive for the total energy consumption.
Under the alternative hypothesis, however, the house temperature may have different effects on the energy consumption at different times of the day. For example, periods of extreme heat (e.g., late‑afternoon peaks) disproportionately drive air‑conditioning demand compared with cooler morning hours, and many households switch off cooling during overnight sleep. A single constant coefficient $\gamma$ would conceal these diurnal fluctuations, whereas a flexible nonparametric function $\beta_1(t)$ can capture how temperature sensitivity of energy use evolves throughout the day.

\noindent\textbf{Example 2 (EEG–behavior latency):}
Consider the EEG dataset of \cite{hinss2023open}, collected during the Psychomotor Vigilance Test (PVT), which is a popular reaction-time task in cognitive neuroscience for assessing sustained attention and alertness \citep{basner2021response}. For each trial $i$ and electrode $j$, the neural voltage trajectory $X_{ij}(t)$ is recorded alongside a behavioral outcome $Y_i$ (i.e., reaction time). 
Empirical studies have shown that behavioral responses often correlate most strongly with neural activity confined to specific post‑stimulus latency windows \citep{luck2014introduction}.
A parsimonious parametric model assumes that only activity near a single latency $\tau$ drives behavior, by centering a Gaussian kernel:
\begin{align*}
    \beta_j(t)=A_j \exp \left\{-\frac{(t-\tau)^{2}}{2 \kappa^{2}}\right\}
\end{align*}
so that the predicted effect peaks at $t=\tau$ and decays symmetrically. By contrast, a fully nonparametric specification allows $\beta_j(t)$ to flexibly adapt to multiple sensitive windows, asymmetric profiles, or other complex temporal dynamics that a single Gaussian shape function cannot capture.

These examples correspond to three practical scenarios: (i) an insignificant functional covariate (zero coefficient), (ii) a covariate with a simple constant or parametric effect, and (iii) a covariate with a genuinely complex and nonparametric effect.  Hence, it is of interest to develop a unified approach that simultaneously identifies which scenario applies to each predictor and estimates the corresponding coefficient functions under the latter two alternatives.
This idea is analogous to the linear and nonlinear discovery (LAND) approach introduced in \cite{zhang2011linear} for standard nonparametric regression, which differentiates among zero, linear, and nonlinear effects for scalar covariates.

In this paper, we propose \textbf{MoFI-FLR} (abbreviated \textbf{MoFI}): a novel two‐stage procedure for high‐dimensional functional linear regression with model‐form identification. 
In the first stage, we apply a functional elastic‐net penalty \citep{guo2024rkhs} to screen out all insignificant functional covariates, thereby reducing the model to a subset of truly predictive covariates. 
In the second stage, each retained (non-zero) coefficient function is decomposed into a simple, finite‐dimensional parametric component and a complementary, infinite‐dimensional nonparametric component. To achieve this, we embed the coefficient functions in a reproducing kernel Hilbert space (RKHS) and perform an orthogonal decomposition of the RKHS into its null subspace (capturing simple component) and its complementary subspace (capturing complementary component), see \cite{hsing2015theoretical} for a comprehensive account of the RKHS theory. Our approach applies regularization exclusively to the complementary subspace, thereby automatically shrinking residual nonparametric variation toward zero. This selective penalization effectively separates simple parametric effects, which contain only the simple component, from complex nonparametric effects, which also contain a complementary component.

\vspace{0.5em}
Our \textbf{main contributions} are threefold:

\begin{enumerate}
  \item We introduce a novel two‐stage estimation framework for high‐dimensional functional linear regression that automatically identifies and isolates simple (parametric), complex (nonparametric), and null effects.
  \item We establish rigorous theoretical guarantees for form‐identification consistency, proving that our procedure correctly recovers the true parametric versus nonparametric structure with high probability.
  \item We develop a computationally efficient algorithm based on block‐coordinate updates and low‐rank eigensystem truncation, which scales readily to large numbers of functional covariates and high‐resolution trajectories.
\end{enumerate}

The remainder of the paper is organized as follows. In Section \ref{sec:method}, we introduce the problem setup and outline our two‐step estimation procedure for MoFI-FLR. Section \ref{sec:theory} develops non‑asymptotic theory and establishes sufficient conditions for selection consistency. In Section \ref{sec:algorithm}, we present a computationally efficient algorithm for estimating the coefficient functions and distinguishing different signal types. Section \ref{sec:simulations} reports the results of our simulation studies, while Section \ref{sec:application} applies MoFI-FLR to task EEG data from the Psychomotor Vigilance Test (PVT) in the COG‑BCI database. Finally, Section \ref{sec:discussion} summarizes our contributions and highlights key findings. The proofs of the main theorems can be found in the Appendix.

\section{Methodology}
\label{sec:method}

Without loss of generality, we assume that all the domains $\mathcal{T}_j=[0,1]$, with the notion that any function on a compact interval can be transformed easily to have $[0,1]$ as its domain. 
Let $\mathbb{L}_2[0,1]$ be the $L_2$-space of square-integrable, measurable functions on $[0,1]$, equipped with the inner product $\langle f, g\rangle_{2} = \int_0^1 f(t)g(t)dt$ and functional norm $\|f\|_2=\langle f, f\rangle_2^{1/2}$, for any $f, g \in \mathbb{L}_2[0,1]$.
We will also be concerned with the $p$-fold product space of $\mathbb{L}_2^p[0,1]$ containing
elements $\boldsymbol{f}=(f_1,\ldots, f_p)^\top$ with each $f_j\in \mathbb{L}_2[0,1]$, $\|\boldsymbol{f}\|_2 \equiv (\sum_{j=1}^p \|f_j\|_2^2)^{1/2} <\infty $ and inner product 
$\langle \boldsymbol{f}, \boldsymbol{g}\rangle_{2} \equiv \sum_{j=1}^p \langle f_j, g_j\rangle_{2}$ for 
$\boldsymbol{f}=(f_1,\ldots, f_p)^\top, \boldsymbol{g} = (g_1, \ldots,g_p)^\top$. Let $\otimes$ be
the outer product associated with either inner product such that $f\otimes g$ defines an operator $(f\otimes g) h = f \langle g, h\rangle_2$.
Then, we can rewrite \eqref{eq:aim3a}: 
\begin{align} \label{e:model} 
    Y_i =  \sum_{j=1}^p \langle X_{ij},  \beta_{j} \rangle_{2} + \varepsilon_i,
\end{align}
where the functional predictors $X_{ij}(\cdot)$ are random elements in $\bbL_2[0,1]$ and $\varepsilon_i$ are iid zero-mean random errors with variance $\sigma^2$.
We further assume that $\beta_{j}(\cdot) \in \mathbb{H}_j := \mathbb{H}(K_j)$, which is the reproducing kernel Hilbert space (RKHS) with kernel $K_j$ \citep{wahba1990spline}. Recall that a real, symmetric, square-integrable, and nonnegative definite function $K(\cdot, \cdot)$ on $[0,1]^2$ is called a reproducing kernel (RK) for a Hilbert space of functions $\mathbb{H}(K)$ on $[0,1]$ if $K(\cdot, t) \in \mathbb{H}(K)$ for any $t\in [0,1]$ and $\mathbb{H}(K)$ is equipped with the inner product such that $\langle \beta, K(\cdot, t)\rangle_{\mathbb{H}(K)} = \beta(t)$ for any $\beta\in \mathbb{H}(K)$ and any $t\in [0,1]$. 

To decompose the functional predictor, we express $\beta_j(t)$ as the sum of two orthogonal components:
\begin{align}
\label{equ:coef_decompose}
    \beta_j(t) = \beta_j^{(0)}(t) + \beta_j^{(1)}(t),
\end{align}
where $\beta_j^{(0)}(\cdot)$ is the simple component, belonging to a finite-dimensional RKHS of interest ($\mathbb{H}_{0j} := \mathbb{H}(K_{0j})$), which is also a subspace of $\mathbb{H}_j$, and $\beta_j^{(1)}(\cdot)$ is the complementary component, belonging to the complementary space ($\mathbb{H}_{1j} := \mathbb{H}(K_{1j})$). By definition, we have $\mathbb{H}_j = \mathbb{H}_{0j} \oplus \mathbb{H}_{1j}$. As a result,
the two components in \eqref{equ:coef_decompose} are orthogonal, i.e.,
$\langle \beta_j^{(0)}, \beta_j^{(1)} \rangle_{\mathbb{H}_j} = 0.$ 
Moreover, the kernel associated with $\mathbb{H}_j$ satisfies
$K_j = K_{0j} + K_{1j}$.

Under this representation, the three scenarios for $\beta_j(t)$ respectively corresponds to: 
\begin{enumerate}
    \item[(i)] Null signal: \(\beta_j(t)\equiv 0\);
    \item[(ii)] Simple (structured) signal: \(\beta_j^{(0)}(t)\not\equiv 0\) and \(\beta_j^{(1)}(t)\equiv 0\);
    \item[(iii)] Complex (unstructured) signal: \(\beta_j^{(1)}(t)\not\equiv 0\).
\end{enumerate}

The interpretation of the simple component depends on the choice of \(\mathbb{H}_{0j}\). For example, if \(\mathbb{H}_{0j}\) is spanned by the constant and linear functions, then the simple component corresponds to a linear trend, while the complementary component represents departures from linearity. In applications such as resting-state EEG analysis, \(\mathbb{H}_{0j}\) may instead be defined using basis functions within a specified frequency range, so that the simple component captures signals in the corresponding frequency band. This user-defined construction allows the notion of a simple component to be tailored to the scientific application.

Next, we propose a two-step procedure for identifying the functional form of coefficients. In the first step, we apply the functional elastic-net method \citep{guo2024rkhs} for variable selection, and we remove any predictors with null coefficients (i.e., $\beta_j(t) \equiv 0$). In the second step, we examine whether $\beta_j^{(1)}(t) \equiv 0$ on the remaining nonzero coefficients to identify the simple and complex signals.

\subsection{Step-One (Variable Selection)}

To promote sparsity in the vector of functional coefficients $\bdbeta = (\beta_1, \ldots, \beta_p)^\top$, where $\bdbeta \in \mathbb{H} := \bigotimes_{j=1}^p \mathbb{H}_j$ (the direct product of RKHSs; see \cite{hsing2015theoretical}), we employ the functional elastic-net method \citep{guo2024rkhs} for estimation:
\begin{align}
\label{equ:mini0}
    \wh \bdbeta=\argmin_{\bdbeta \in \mathbb{H}} \left\{ \frac{1}{2n} \sum_{i=1}^n \left(Y_i -  
\sum_{j=1}^p \langle X_{ij},  \beta_j \rangle_{2}  \right)^2 
+  
    \sum_{j=1}^p \mathrm{Pen}(\beta_j; \bdlambda)\right\} 
\end{align}
where $\mathrm{Pen}(\beta_j; \bdlambda)$ 
is the functional elastic-net penalty to be specified below with  $\bdlambda$ denoting a vector of tuning parameters. 
Instead of directly optimizing $\beta_j$, it has been shown that optimizing a transformed version of $\beta_j$ is computationally more efficient. Specifically, denote by $\mathscr{K}$ the integral operator of kernel $K$: $(\mathscr{K} f)(\cdot)= \int_0^1 K(s, \cdot) f(s) ds, \ f\in \mathbb{L}_2[0,1]$. Suppose $K$ has a spectral decomposition $K(s,t)=\sum_{k=1}^\infty \theta_{k}\varphi_{k}(s) \varphi_{k}(t)$. Then its square root is defined as 
$K^{1/2}(s,t)=\sum_{k=1}^\infty \theta_{k}^{1/2} \varphi_{k}(s) \varphi_{k} (t)$, and  
$\mathscr{K}^{1/2}$ is the associated square-root integral operator. 
By \cite{CaiYuan2012}, $\mathscr{K}^{1/2}(\mathbb{L}_2[0,1]) = \mathbb{H}(K)$. Therefore, for all $\beta \in  \mathbb{H}(K)$, there exists a unique 
$f \in \mathbb{L}_2[0,1]$ such that $\beta = \mathscr{K}^{1/2}f$ with $\| \beta \|_{\mathbb{H}(K)} = \| f \|_{2}$. Without causing any confusion, we use $\|\cdot\|_2$ to denote the norm of $\mathbb{L}_2$ functions or vectors of $\mathbb{L}_2$ functions as well as the Euclidean norm in $\mathbb{R}^p$.

With the above $\bbL_2$ representation $\boldsymbol{f}$ of $\bdbeta$, the loss function in (\ref{equ:mini0}) can be rewritten as
\begin{align} \label{equ:mini1}
     \frac{1}{2n} \sum_{i=1}^n \left(Y_i -  
\sum_{j=1}^p \langle \wt{X}_{ij},  f_j \rangle_{2}  \right)^2 
+  \sum_{j=1}^p \mathrm{Pen}(f_j; \bdlambda),
\end{align}
where $\wt{X}_{ij} = \mathscr{K}_j^{1/2} X_{ij}$. The following functional elastic-net penalty \citep{guo2024rkhs} is used
\begin{align}
\label{equ:fenet_pen}
	\mathrm{Pen}(f_j; \lambda_1, \lambda_2, \Psi_j)= \lambda_1  \|\Psi_j f_j \|_{2} + \frac{\lambda_2}{2} \|f_j\|_{2}^2,  \quad \lambda_1,\lambda_2 >0,
\end{align}
In practice, we specify $\Psi_j = (\Tnjj + \theta \mathscr{I})^{1/2}$, where $\Tnjj$ is the empirical covariance of $(\wt X_{1j}, \cdots, \wt X_{nj})^{\top}$, $\mathscr{I}$ is the identity operator that maps a function to itself, and we let $\theta>0$.

To better understand \eqref{equ:fenet_pen}, note that the first term primarily encourages sparsity, whereas the second term serves as a ridge-type smoothness penalty. Under the above choice of \(\Psi_j\), the first term can be written as
\begin{align}
\|\Psi_j f_j\|_2
&=
\left(
\frac{1}{n}\sum_{i=1}^{n}\langle \wt X_{ij}, f_j\rangle_2^{2}
+
\theta \|f_j\|_2^2
\right)^{1/2}.
\label{equ:pen_decomp}
\end{align}
The first part inside the square root in \eqref{equ:pen_decomp}, namely
$n^{-1}\sum_{i=1}^{n}\langle \wt X_{ij}, f_j\rangle_2^{2},$
measures the predictive signal strength of the \(j\)-th functional predictor and is commonly used as a penalty term in sparse additive models, including high-dimensional FLRs \citep{ravikumar2009sparse, XueYao2021}.
The second part in \eqref{equ:pen_decomp} involving \(\theta\) regularizes the spectrum of \(\mathscr T_n^{(j,j)}\) by shifting its eigenvalues away from zero. Consequently, taking \(\theta>0\) ensures that the sparse penalty remains well defined even when the \(j\)-th predictor has only a weak association with \(Y\).

Next, let $\widehat{\boldsymbol{f}}=(\widehat{f}_1, \ldots, \widehat{f}_p)^\top$ denote the functional elastic-net estimate of the coefficients. We define the estimated signal set as $\widehat{\mathcal{S}} = \left\{ j \in \{1,\dots,p\} : \widehat{f}_j \not\equiv 0 \right\}$,
and subsequently remove all functional predictors corresponding to indices not in $\widehat{\mathcal{S}}$.

\subsection{Step-Two (Model Form Identification)}

In this step, we restrict our attention to the predictors in the estimated signal set $\wh{\calS}$.
Recall that $\beta_j^{(0)} \in \mathbb{H}_{0j}$
Assume that $\mathbb{H}_{0j}$ is an $M_0$-dimension RKHS. Then, there exists an orthogonal basis $\{\phi_{jm}\}_{m=1}^{M_0}$ in $\mathbb{H}_{0j}$ and a sequence $\nu_{j1} \ge \dots \ge \nu_{jM_0} > 0$ such that
\begin{align*}
    K_{0j}(s,t) = \sum_{m=1}^{M_0} \nu_{jm} \phi_{jm}(s) \phi_{jm}(t).
\end{align*}
Consequently, we can represent 
\begin{align*}
    \beta_{j}^{(0)}(t) = \sum_{m=1}^{M_0} a_{jm} \phi_{jm}(t)
\end{align*}
for some $\mathbf{a}_j = (a_{j1}, \dots, a_{jM_0})^{\top}$.

\begin{remark}
In practice, $K_{0j}$ is often not given. Instead, we assume that the null–space component $\beta_j^{(0)}$ lies in the span of the (not necessarily orthogonal) basis functions $\{\psi_{jm}\}_{m=1}^{M_0}\subset\mathbb{H}(K_j)$,
which we collect into the vector $\boldsymbol\psi_j(t)=\bigl(\psi_{j1}(t),\dots,\psi_{jM_0}(t)\bigr)^\top$.
To construct the reproducing kernel $K_{0j}$, define the $M_0\times M_0$ Gram matrix $G_j = \Bigl[\langle\psi_{jr_1},\,\psi_{jr_2}\rangle_{\mathbb{H}_j}\Bigr]_{r_1,r_2=1}^{M_0}$.
Then $K_{0j}$ admits the representation
\begin{align*}
    K_{0j}(s,t) = \boldsymbol\psi_j(s)^\top\,G_j^{-1}\,\boldsymbol\psi_j(t).
\end{align*}
One readily checks this by noting that $\langle \beta_j^{(0)}, \ K_{0j}(\cdot, t) \rangle_{\mathbb{H}_j}  =
    \beta_j^{(0)}(t)$ for any $\beta_j^{(0)}(t) = \mathbf{a}_j^{\top} \ \boldsymbol\psi_j(t)$ with $\mathbf{a}_j \in \mathbb{R}^{M_0}$.

\end{remark}

Similar to Step-One, for all $\beta_j^{(1)} \in  \mathbb{H}_{1j}$, there exists a unique 
${f}_j^{(1)} \in \mathbb{L}_2[0,1] / \mathrm{Ker}(\mathscr{K}_{1j}^{1/2})$ such that $\beta^{(1)}_j = \mathscr{K}_{1j}^{1/2} f^{(1)}_j$ with $\| \beta^{(1)}_j \|_{\mathbb{H}_{1j}} = \| f^{(1)}_j \|_{2}$. 
Define $\mathbf{Z}_{ij} = (Z_{ij1}, \dots, Z_{ijM_0})^{\top}$, where $Z_{ijm} = \langle X_{ij}, \phi_{jm} \rangle_2$. Let $\wt{X}_{ij}^{(1)} = \mathscr{K}_{1j}^{1/2} X_{ij}$.
We estimate $\{(\mathbf{a}_j, f^{(1)}_j)\}_{j \in \wh{\cal S}}$ by minimizing
\begin{align} \label{equ:mini2}
     \frac{1}{2n} \sum_{i=1}^n \left( Y_i - \sum_{j \in \wh{\cal S}} \mathbf{Z}_{ij}^{\top} \mathbf{a}_j - 
\sum_{j \in \wh{\cal S}}  \left\langle \wt{X}_{ij}^{(1)},  f^{(1)}_j \right\rangle_{2} \right)^2 
+  \sum_{j \in \wh{\cal S}} \mathrm{Pen}(f^{(1)}_j; \widebar\lambda_1, \widebar\lambda_2, \Psi_j^{(1)}),
\end{align}
where
\begin{align*}
	\mathrm{Pen}(f_j^{(1)}; \bar\lambda_1, \bar\lambda_2, \Psi_j^{(1)})= \bar\lambda_1  \left\|\Psi_j^{(1)} f_j^{(1)} \right\|_{2} + \frac{\bar\lambda_2}{2} \left\|f_j^{(1)} \right\|_{2}^2,  \quad \bar\lambda_1, \bar\lambda_2 >0.
\end{align*}
In practice, we let $\Psi_j^{(1)} = (\Tlnjj + \theta \mathscr{I})^{1/2}$, where $\Tlnjj$ denotes the empirical covariance matrix of
$\bigl( \widetilde{X}_{1j}^{(1)}, \dots, \widetilde{X}_{nj}^{(1)} \bigr)^{\top}$.
Then the simple (structured) signal set is estimated by 
\begin{align*}
    \widehat{\mathcal{S}}_0 = \left\{ j \in \wh{\calS} : \ \widehat{f}_j^{(1)} \equiv 0 \right\};
\end{align*}
and the complex (unstructured) signal set is estimated by 
\begin{align*}
    \widehat{\mathcal{S}}_1 = \left\{ j \in \wh{\calS} : \  \widehat{f}_j^{(1)} \not\equiv 0 \right\}.
\end{align*}
Clearly, $\widehat{\mathcal{S}} = \widehat{\mathcal{S}}_0 \bigcup \widehat{\mathcal{S}}_1$.

\noindent
\begin{remark}
Define $\wt{X}_{ij}^{(0)}=\mathscr{K}_{0j}^{1/2}X_{ij}$. As for $f_j^{(1)}$, there exists a unique 
${f}_j^{(0)}$ such that $\beta^{(0)}_j = \mathscr{K}_{0j}^{1/2} f^{(0)}_j$ with $\| \beta^{(0)}_j \|_{\mathbb{H}_{0j}} = \| f^{(0)}_j \|_{2}$. 
Because
\begin{align*}
   \left\langle X_{ij}, \beta_j \right\rangle_2 =
   \left\langle X_{ij}, \beta_j^{(0)} \right\rangle_2 + \left\langle X_{ij}, \beta_j^{(1)} \right\rangle_2 = \left\langle \wt X_{ij}^{(0)}, f_j^{(0)} \right\rangle_2 + \left\langle \wt X_{ij}^{(1)}, f_j^{(1)} \right\rangle_2,
\end{align*}
we may replace $\mathbf{Z}_{ij}^{\top}\mathbf{a}_j$ in \eqref{equ:mini2} by
$\langle \wt{X}_{ij}^{(0)}, f_j^{(0)}\rangle_2$. 
This further emphasizes that the predictor can be decomposed into structured and complementary components, while regularization is applied only to the coefficient corresponding to the complementary component.

\end{remark}

\section{Theory}
\label{sec:theory}

\subsection{Notation and Preliminaries}

We first introduce some notation to be used in this section. 
Let $\mathbb{H}_1$ and $\mathbb{H}_2$ be two Hilbert spaces and $\mathscr{F}: \mathbb{H}_1 \to \mathbb{H}_2$ be a compact linear operator mapping from $\mathbb{H}_1$ to $\mathbb{H}_2$. Then the $\mathbb{L}_2$ operator norm is defined as $\|\mathscr{F}\|_2= \sup_{f\in \mathbb{H}_1} \|\mathscr{F} f\|_2/ \|f\|_2$ which is the maximum singular value of $\mathscr{F}$; if $\mathbb{H}_1=\mathbb{H}_2$ and $\mathscr{F}$ is self-adjoint, the trace $\tr(\mathscr{F})$ is defined as the sum of all eigenvalues of $\mathscr{F}$.
For any $\boldsymbol{f} =(f_{1}, \ldots, f_p)^\top \in \mathbb{L}_2^p[0,1]$,  $\| \boldsymbol{f} \|_{\infty} = \max_j \| f_j \|_{2}$; 
for any $r \times s$ operator-valued matrix $\mathscr{F}=( \mathscr{F}_{ij})_{i,j=1}^{r,s}$, where each $\mathscr{F}_{ij}$ maps from $\mathbb{L}_2[0,1]$ to $\mathbb{L}_2[0,1]$, 
 define the norm $\vertiii{\mathscr{F}}_{a,b} = \sup_{\| \boldsymbol{f} \|_{a} \le 1} \|\mathscr{F} \boldsymbol{f} \|_{b}$ for $a, b \in \{2, \infty\}$. 
For any index sets ${\calS}_1$ and ${\calS}_2$, $\mathscr{F}^{({\calS}_1, {\calS}_2)}$ is the submatrix of $\mathscr{F}$ with rows in ${\calS}_1$ and columns in ${\calS}_2$.

Next, we assume that $\BX_i=(X_{i1},\ldots, X_{ip})^\top \in \mathbb{L}_2^p[0,1]$, $i=1, \ldots, n$, are iid zero-mean Gaussian random elements.
Define the covariance operator
\begin{align} \label{e:cov_op}
    \T = \E \left\{ (\wt{X}_{i1},\ldots, \wt{X}_{ip})^\top \otimes (\wt{X}_{i1},\ldots, \wt{X}_{ip}) \right\}.
\end{align}
It is convenient to view $\T$ as a $p\times p$ operator-valued matrix 
$\{\T^{(j,j')}\}$ where $\T^{(j,j')}=\E (\wt{X}_{ij}\otimes \wt{X}_{ij'})$ is the cross covariance operators of $\wt{X}_{ij}$ and $\wt{X}_{ij'}$. 
Denote $\T^{(j,j)}_\lambda=\T^{(j,j)} + \lambda \scrI$ for any $\lambda>0$ where $\mathscr{I}$ is the identity operator. 
Let $\QSS=\mathrm{diag}\{\mathscr{T}^{(j,j)}, j \in \calS\}$ be the operator-valued matrix that only contains the diagonal terms of $\TSS$. Let $\TSS_{\lambda} = \TSS + \lambda \mathscr{I}$, $\QSS_{\lambda} = \QSS + \lambda \mathscr{I}$. 
Additionally, define the underlying true functional coefficient as $\boldsymbol{f}_0=(f_{01}, \ldots, f_{0p})^\top$.
Define the true signal set $\calS = \{j \in  \{1, \dots, p\}: f_{0j} \not\equiv 0 \}$, the true simple signal set $\calS_0 = \{j \in  \calS: \ f_{0j}^{(1)} \equiv 0 \}$, and the true complex signal set $\calS_1 = \{j \in  \calS:  \ f_{0j}^{(1)} \not\equiv 0 \}$.
Let the cardinalities of $\calS$, $\calS_0$, and $\calS_1$ be respectively $q$, $q_0$ and $q_1$. Clearly, $q_0 + q_1 = q$.

Define the $n \times qM_0$ matrix $\mathbf{Z}_{\calS} = (\mathbf{Z}_{ij}^{\top} )_{1\le i \le n}^{j \in \calS}$. Let  
$H_{\calS} = I_n - \bZ_{\calS} (\bZ_{\calS}^{\top} \bZ_{\calS} )^{-1} \bZ_{\calS}^{\top}$
be the orthogonal projection onto the complement of the column space of $\bZ_{\calS}$.
For any index set $\cal W$ (in particular ${\cal W} \in \{ {\cal S}, {\cal S}_0, {\cal S}_1 \}$), let $\wt{X}_{\cal W}^{(1)} = ( \wt{X}_{ij}^{(1)})_{1\le i \le n}^{j \in \cal W}$.
Define
$\boldsymbol{U}_{\cal W} = H_{\calS} \wt{X}_{\cal W}^{(1)}$,
\begin{align*}
    \mathscr{A}_n^{({\cal W}_1, {\cal W}_2)} = n^{-1} \boldsymbol{U}_{{\cal W}_1}^{\top} \otimes \boldsymbol{U}_{{\cal W}_2} \quad \mbox{and} \quad
    \mathscr{A}^{({\cal W}_1, {\cal W}_2)} = \mathbb{E} \left(\mathscr{A}_n^{({\cal W}_1, {\cal W}_2)} \right).
\end{align*}
Let $\mathscr{B}^{({\cal S}_1, {\cal S}_1)}=\mathrm{diag}\{\mathscr{A}^{(j,j)}, j \in \calS_1\}$ be the operator-valued matrix that only contains the diagonal terms of $\mathscr{A}^{({\cal S}_1, {\cal S}_1)}$, $\mathscr{B}_{\lambda}^{({\cal S}_1, {\cal S}_1)} = \mathscr{B}^{({\cal S}_1, {\cal S}_1)} + \lambda \mathscr{I}$.

\subsection{Selection Consistency}

We next establish the selection consistency for model-form identification:
\begin{align}
    \Pr\bigl(\widehat{\mathcal S}_0 = \mathcal S_0,\ \widehat{\mathcal S}_1 = \mathcal S_1\bigr) \longrightarrow 1.
    \label{equ:prob_form}
\end{align}
Observe that
\begin{align*}
    \Pr\bigl(\widehat{\mathcal S}_0 = \mathcal S_0,\ \widehat{\mathcal S}_1 = \mathcal S_1\bigr)
= \Pr\bigl(\widehat{\mathcal S}_0 = \mathcal S_0 \mid \widehat{\mathcal S} = \mathcal S\bigr)\,
  \Pr\bigl(\widehat{\mathcal S} = \mathcal S\bigr).
\end{align*}
Thus, model-form identification consistency decomposes into two steps:

\noindent (1) Variable‐selection consistency: by \cite{guo2024rkhs}, with proper conditions, we have
\begin{align*}
    \Pr(\widehat{\mathcal S} = \mathcal S) > 1 - P_1, \quad P_1 = \exp \left(-D \lambda_{2}^{2} n/ q \right)
\end{align*}
for some $D>0$. 

\noindent (2) Conditional form‐selection consistency: Conditioning on $\widehat{\mathcal S} = \mathcal S$, we want to show 
\begin{align}
    \Pr\bigl(\widehat\calS_0 = \calS_0 \mid \widehat\calS = \calS\bigr) = \Pr\bigl(\widehat\calS_1 = \calS_1 \mid \widehat\calS = \calS\bigr) > 1 - P_2,
    \label{equ:ineq_s1}
\end{align}
with $P_2$ shrinking to zero as $n \rightarrow \infty$.
It follows that the left-hand side of \eqref{equ:prob_form} is bounded below by $1 - P_1 - P_2$, which converges to one as $n \rightarrow \infty$
Hence, the main technical challenge is to establish the conditional bound in \eqref{equ:ineq_s1}.

We need the following conditions to proceed. 

\begin{condition}
For $j=1,\dots,p$, $\Psi_j$ is a self-adjoint operator such that $\Psi_j f \in \mathbb{M}_{nj}$ for all $f \in \mathbb{M}_{nj}$, where $\mathbb{M}_{nj}$ is the minimal space spanned by $\wt{X}_{1j}, \dots, \wt{X}_{nj}$.
Assume that there exist positive constants $0<C_{\min} <C_{\max}< \infty$ such that, uniformly for all $j$,  the eigenvalues of $\Psi_j$ are in the interval $[C_{\min},C_{\max}]$.
\label{ass:a1}
\end{condition}

\begin{condition}
Each $\mathscr{T}^{(j,j)}$ is standardized such that $\|\scrT^{(j,j)}\|_2=1$, 
with its trace uniformly bounded by a finite constant $\tau$, i.e.,
$\sup_{j \in \{1, \dots, p\}} \{ \tr(\mathscr{T}^{(j,j)}) \}  \le \tau$.
\label{ass:a2}
\end{condition}
\begin{condition}
Define
$\varkappa(\lambda_2) = \vertiii{ \TSS (\TSS_{\lambda_2})^{-1} }_{\infty, \infty}$.
Assume that for some $\gamma\in (0,1]$, we have
$\varkappa(\lambda_2) \cdot \vertiii{ \TScS (\TSS)^{-} }_{\infty, \infty} \le (C_{\min} / C_{\max}) (1 - \gamma)$,
where $(\TSS)^{-}$ is the Moore-Penrose generalized inverse of $\TSS$.
\label{ass:a3}
\end{condition}
\begin{condition}
$\aleph(\lambda_2)= \vertiii{ \left( \mathscr{T}^{(\calS, \calS)} - \mathscr{Q}^{(\calS, \calS)} \right)  \big(\mathscr{Q}^{(\calS, \calS)}_{\lambda_2}\big)^{-1} }_{\infty, \infty} < 1$.
\label{ass:a4}
\end{condition}
\begin{condition}
Define $G =  \left\{ 8 + 4 (1 - \aleph(\lambda_2))^{-1} \right\} \left( C_{\max}  \lambda_1\lambda_2^{-1/2} + 2 \lambda_2^{1/2} \right)$, we assume that
$\min_{j \in \calS} \left\| (\Tjj)^{1/2} f_{0j} \right\|_2 > G$.
\label{ass:a5}
\end{condition}

\begin{condition}
For $j = 1, \dots, p$, $\Psi_j^{(1)}$ is a self-adjoint operator such that $\Psi_j^{(1)} f \in \mathbb{M}_{nj}^{(1)}$ for all $f \in \mathbb{M}_{nj}^{(1)}$, where $\mathbb{M}_{nj}^{(1)}$ is the minimal space spanned by $\wt{X}_{1j}^{(1)}, \dots, \wt{X}_{nj}^{(1)}$. Furthermore, for all $j$,  the eigenvalues of $\Psi_j^{(1)}$ are in the interval $[C_{\min},C_{\max}]$.
\label{ass:a6}
\end{condition}

\begin{condition}
Define
$\bar\varkappa(\bar\lambda_2) = \vertiii{ \mathscr{A}^{({\cal S}_1, {\cal S}_1)} (\mathscr{A}^{({\cal S}_1, {\cal S}_1)}_{\bar\lambda_2})^{-1} }_{\infty, \infty}$.
Assume that for some $\bar\gamma\in (0,1]$, we have
$\bar\varkappa(\bar\lambda_2) \cdot \vertiii{ \mathscr{A}^{({\cal S}_0, {\cal S}_1)} ( \mathscr{A}^{({\cal S}_1, {\cal S}_1)} )^{-} }_{\infty, \infty} \le (C_{\min} / C_{\max}) (1 - \bar\gamma)$.
\label{ass:a7}
\end{condition}

\begin{condition}
$\bar\aleph(\bar\lambda_2):= \vertiii{ \left( \mathscr{A}^{({\cal S}_1, {\cal S}_1)} - \mathscr{B}^{({\cal S}_1, {\cal S}_1)} \right)  \big( \mathscr{B}_{\bar\lambda_2}^{({\cal S}_1, {\cal S}_1)} \big)^{-1} }_{\infty, \infty} < 1$.
\label{ass:a8}
\end{condition}

\begin{condition}
Define
$\bar G = \left\{ 8 + 4 (1 - \bar\aleph( \bar\lambda_2))^{-1} \right\} \left( C_{\max}  \bar\lambda_1 \bar\lambda_2^{-1/2} + 2 \bar\lambda_2^{1/2} \right)$, we assume that
$\min_{j \in {\calS}_1 } \left\| (\mathscr{A}^{(j,j)})^{1/2} f_{0j}^{(1)} \right\|_2 > \bar{G}$.
\label{ass:a9}
\end{condition}

\noindent
\begin{remark}
\noindent \textbf{(i):}
Condition \ref{ass:a1} specifies the admissible class of penalty operators. Following \cite{guo2024rkhs}, we adopt the practical choice
$\Psi_j = (\Tnjj + \theta \mathscr{I})^{1/2}$ with $\theta>0$,
which satisfies Condition \ref{ass:a1} and penalizes each predictor according to its predictive power. \\
\noindent \textbf{(ii):}
Condition \ref{ass:a2} places a mild constraint on the decay rate of the eigenvalues for $\mathscr{T}^{(j,j)}$, 
Condition \ref{ass:a2} also implies $\sup_{j \in \{1, \dots, p\}} \{ \tr(\mathscr{T}_1^{(j,j)}) \}  \le \tau$, where $\mathscr{T}_1^{(j,j)} = \mathbb{E}(\wt{X}_{1j}^{(1)} \otimes \wt{X}_{1j}^{(1)})$.\\
\noindent \textbf{(iii):}
Condition \ref{ass:a3} controls the correlation between functional predictors in the true signal set $\calS$ and those in the non-signal set $\calS^c$. This assumption is related to the ``irrepresentable condition'' on model selection consistency of the classical lasso \citep{zhao2006model, wainwright2009sharp}. When the predictors in $\calS$ and in $\calS^c$ are uncorrelated, the condition holds trivially. 
By contrast, if predictors in \(\calS^c\) are strongly correlated with those in \(\calS\), then the non-signal covariates may mimic the signal variables, thereby making consistent separation and identification much more challenging.

\noindent \textbf{(iv):}
Condition \ref{ass:a4} puts constraints on the correlations between the predictors in the true signal set $\calS$, so that none of the true predictors can be represented by other predictors in $\calS$. It trivially holds when the predictors in $\calS$ are uncorrelated. \\
\noindent \textbf{(v):}
Condition~\ref{ass:a5} specifies a lower bound $G$ on the minimum detectable signal strength. In particular, the signal strength of the \(j\)-th covariate is quantified by $\left\| (\T_{jj})^{1/2} f_{0j} \right\|_2^2 = \mathrm{E}\bigl(\langle X_{ij}, \beta_{0j}\rangle_2^2\bigr)$
and $G$ is assumed to lower bound the smallest nonzero signal. If a signal falls below this threshold, it may become indistinguishable from noise, causing variable selection to fail. \\
\noindent \textbf{(vi):} 
Conditions~\ref{ass:a2}--\ref{ass:a5} ensure variable-selection consistency by imposing constraints on the entire signal set $\calS$ and on the relationship between $\calS$ and $\calS^c$. By contrast, Conditions~\ref{ass:a7}--\ref{ass:a9} guarantee conditional form-identification consistency by imposing analogous constraints on the complex-signal set $\calS_1$ and on the relationship between $\calS_1$ and $\calS_0$. Thus, the latter conditions parallel the former in both structure and purpose. \\
\noindent \textbf{(vii):} In Appendix A, we provide an illustrative example in which the functional predictors follow a partially separable covariance structure \citep{zapata2022partial} with an AR(1)-type dependence across predictors. We examine the validity of our technical assumptions under this setting, particularly Conditions 2-5 and 7-8. Specifically, we show that Conditions~3 and~4 hold under mild correlation, for example, when the autoregressive correlation is smaller than \( 1/3 \) and \(\gamma\) is sufficiently small.
We also show that the operator $\mathscr{A}^{ ( {\cal W}_1, {\cal W}_2 )}$ appearing in Conditions~7 and~8 behaves analogously to the covariance operator of \(\wt X_{ij}^{(1)}\), which suggests that these conditions are mild, especially when Conditions~3 and~4 are assumed to hold.

\end{remark}

The following proposition establishes that the minimization problems for both steps are indeed well defined and any minimizer must be in a finite-dimensional subspace.

\begin{proposition}\label{lemma:solution_form}
Under Condition 1, any minimizer \(\widehat f_j\) \((j=1,\dots,p)\) of \eqref{equ:mini1} in Step-One belongs to the subspace spanned by \(\{\widetilde X_{ij}(\cdot): i=1,\dots,n\}\).

Under Condition 6, any minimizer \(\widehat f_j^{(1)}\) \((j=1,\dots,p)\) of \eqref{equ:mini2} in Step-Two belongs to the subspace spanned by \(\{\widetilde X_{ij}^{(1)}(\cdot): i=1,\dots,n\}\).
\end{proposition}

Proposition \ref{lemma:solution_form} follows from the Riesz representer theorem; its proof is given in the appendix.
The selection consistency for MoFI-FLR is given in the following result.

\begin{theorem} \label{thm:2}
Consider the two-stage problem in \eqref{equ:mini1} and \eqref{equ:mini2}.
Suppose that Conditions \ref{ass:a1}-\ref{ass:a9} hold. 
Suppose the following conditions on $\lambda_1,\lambda_2, \bar\lambda_1, \bar\lambda_2$ hold:
\begin{align}
\begin{split}
        &\lambda_1/\lambda_2 > \left({3\over\gamma} - 2\right) C_{\max}^{-1}, \quad D_{1,1}^* >  \lambda_1 > D_{1,2}^* { \tau^{1/2} (1+\sigma)  \over C_{\min} \gamma} \left\{{\log(p-q) \over n} \right\}^{1/2}, \\
    & D_{2,1}^* > \lambda_2 > D_{2,2}^* \frac{\tau (1+\sigma) (\rho_1 + 1)}{(C_{\min} / C_{\max})^2 \gamma^2}  \max\left\{{q\log(p-q)\over n},\left(q^2\over n \right)^{1/2} \right\}. \\
            &\bar\lambda_1/ \bar\lambda_2 > \left({3\over\bar\gamma} - 2\right) C_{\max}^{-1}, \quad \bar{D}_{1,1}^* >  \bar{\lambda}_1 > \bar{D}_{1,2}^* { \tau^{1/2} (1+\sigma)  \over C_{\min} \bar\gamma} \left\{{\log(q_0) \over n} \right\}^{1/2}, \\
    & \bar{D}_{2,1}^* > \bar\lambda_2 > \bar{D}_{2,2}^* \frac{\tau (1+\sigma) ( \bar{\rho}_1 + 1)}{(C_{\min} / C_{\max})^2 \bar\gamma^2}  \max\left\{{q_1\log(q_0)\over n},\left(q_1^2\over n \right)^{1/2} \right\}.
\end{split}
\label{equ:bar_lambda_12_bound_1}
\end{align}
where $\rho_1$ denotes the largest eigenvalue of $\TSS$, $\bar\rho_1$ denotes the largest eigenvalue of $\All$. $\bar{D}_{1,1}, \bar{D}_{1,2}, \bar{D}_{2,1}, \bar{D}_{2,2}, \bar{D}_{1,1}^*, \bar{D}_{1,2}^*, \bar{D}_{2,1}^*, \bar{D}_{2,2}^*$ are universal constants that do not depend on the model parameters, sample size, or regularization parameters.
Then $\widehat{\calS}_0$ and $\widehat{\calS}_1$ exist uniquely, 
\begin{align}
    \Pr\bigl(\widehat{\mathcal S}_0 = \mathcal S_0,\ \widehat{\mathcal S}_1 = \mathcal S_1\bigr) \ > \  1 -   \exp\left( -D{\lambda_2^2 n \over q } \right) - \exp\left( -\bar{D}{\bar{\lambda}_2^2 n \over q_1 } \right),
\label{equ:vs_consistency_rate1}
\end{align}
where
\begin{align} 
    D = D^* \left\{(C_{\min}/C_{\max}) \gamma \over \tau^{1/2} (\rho_1+1) (\sigma + 1)  \right\}^2
    \quad \mbox{and} \quad
    \bar{D} = \bar{D}^* \left\{(C_{\min}/C_{\max}) \bar{\gamma} \over \tau^{1/2} (\bar\rho_1+1) (\sigma + 1)  \right\}^2,
    \label{equ:D_form}
\end{align}
for some universal constant $D^*$ and $\bar{D}^*$.

\end{theorem}

\begin{remark}
\noindent \textbf{(i):}
Theorem \ref{thm:2} provides a non‑asymptotic guarantee for model‑form identification.  In particular, if
\begin{align}
    \frac{\lambda_2^2\,n}{q}\;\to\;\infty
\quad\text{and}\quad
\frac{\bar\lambda_2^2\,n}{q_1}\;\to\;\infty
\quad\text{as }n\to\infty, \label{equ:lambda_2_cond}
\end{align}
then the right‑hand side of \eqref{equ:vs_consistency_rate1} converges to one.  Moreover, by the bounds in \eqref{equ:bar_lambda_12_bound_1}, one can indeed choose $\lambda_2$ and $\bar\lambda_2$ to satisfy \eqref{equ:lambda_2_cond}.

\noindent 
\textbf{(ii):}
 In \eqref{equ:bar_lambda_12_bound_1}, a comparison of the lower bounds for \(\lambda_1,\lambda_2,\bar\lambda_1,\bar\lambda_2\) suggests that the minimum penalty levels required in Step-Two (model-form identification) are generally smaller than those in Step-One (variable selection). Intuitively, this is because Step-One is performed over all \(p\) covariates, possibly in an ultra-high-dimensional setting with \(p\gg n\), so the penalties must be large enough to control false positives among the \(p-q\) null effects. By contrast, Step-Two is applied after variable selection and therefore operates on the reduced set of \(q\) selected covariates, where typically \(q\ll p\). As a result, a smaller penalty is sufficient to distinguish the \(q_0\) simple effects from the \(q_1\) complex effects. \\
\noindent \textbf{(iii):}
To facilitate the discussion, denote $a_k \asymp b_k$ for two positive sequences $\{a_k\}_{k=1}^\infty$ and $\{b_k\}_{k=1}^\infty$, if $c_1 < a_k / b_k < c_2$ for some $0 < c_1 < c_2 < \infty$ and for all $k$. 
Consider the ultra‑high‑dimensional FLM regime, where 
\begin{align*}
    \log p = O\bigl(n^{1-2\varsigma}\bigr),
\qquad
q \asymp n^{\varsigma},
\qquad
0 < \varsigma < \tfrac14.
\end{align*}
Therefore, setting the tuning parameters \(\lambda_1\), \(\lambda_2\), \(\bar\lambda_1\), and \(\bar\lambda_2\), together with the minimum detectable signal levels \(G\) and \(\bar G\), as
\begin{align*}
    \lambda_1 \asymp \lambda_2 \asymp n^{-\varsigma},
\quad
\bar\lambda_1 \asymp \bar\lambda_2 \asymp n^{-\left(\frac12 - \varsigma\right)},
\quad
G \asymp n^{-\frac{\varsigma}{2}},
\quad
\bar G \asymp n^{-(\frac14 - \frac{\varsigma}{2})},
\end{align*}
satisfies the tuning-parameter requirements in Theorem~\ref{thm:2}.
Consequently, the probability bound in \eqref{equ:vs_consistency_rate1} converges to 1, yielding exact recovery of \(\mathcal S_0\) and \(\mathcal S_1\). We provide the corresponding proof in the Appendix.

\end{remark}

\subsection{Rate of the Excess Prediction Risk}

Following \cite{CaiYuan2012}, we assess the prediction performance of an estimator \(\widehat{\boldsymbol\beta}\) through its excess prediction risk,
\begin{align}
    \mathcal{R}(\widehat{\boldsymbol\beta})
    =
    \mathbb{E}^*\left(
    \sum_{j=1}^p \langle X_{ j}^*, \beta_{0j}-\widehat\beta_j \rangle_2
    \right)^2,
    \label{equ:excess_risk}
\end{align}
where \( \BX^*=(X_{ 1}^*,\dots,X_{ p}^*)\) is an independent copy of \(\BX_{i}=(X_{i1},\ldots, X_{ip})^\top \), and \(\mathbb{E}^*\) denotes expectation with respect to \( \BX^*\).
By Theorem 2 of \cite{guo2024rkhs}, if \(\widehat{\boldsymbol\beta}^{(\mathrm{fenet})}\) denotes the Step-One estimator obtained by minimizing \eqref{equ:mini1}, then, for sufficiently large \(n\), 
\[
\mathcal{R}\!\left(\widehat{\boldsymbol{\beta}}^{(\mathrm{fenet})}\right)
=
O_p\!\left(q\frac{\lambda_1^2}{\lambda_2} \right).
\]

Now let \(\widehat{\boldsymbol\beta}^{(\mathrm{MoFI})}\) denote the Step-Two estimator obtained by minimizing \eqref{equ:mini2}. Owing to technical difficulties, we have not yet been able to establish a comparable prediction risk result in full generality.
Instead, we establish an analogous result under a special case:

\begin{condition}\label{cond:common_rkhs_basis}
Assume that
$\mathbb{H}_{0j}(K_{0j})=\mathbb{H}_0(K_0)$ and 
$\mathbb{H}_{1j}(K_{1j})=\mathbb{H}_1(K_1)$,
so that \(\mathbb{H}_j(K_j)=\mathbb{H}(K)\), \(j=1,\dots,p\). In addition, $\BX_i$ satisfies a partially separable covariance
structure \citep{zapata2022partial} as described in (\ref{eq:partial_separable}) in Appendix A, and that the covariance operator of \(X_{ij}\) and the kernel operator
\(K\) share a common set of eigenfunctions.
\end{condition}

Appendix A provides a practical setting where Condition~\ref{cond:common_rkhs_basis} holds.
Under this special case, the prediction error bound can be derived explicitly.

\begin{corollary}\label{cor:rate_mofi_special}
Under Condition~\ref{cond:common_rkhs_basis} and the conditions of Theorem~\ref{thm:2}, 
\begin{align}
    \mathcal{R}\!\left(\widehat{\boldsymbol{\beta}}^{(\mathrm{MoFI})}\right)
    =
    O_p\!\left\{
\frac{qM_0}{n}
\left(
1+q_1\frac{\bar\lambda_1^2}{\bar\lambda_2^2}
\right)
+
q_1\frac{\bar\lambda_1^2}{\bar\lambda_2}
\right\} \quad \hbox{as \(n \to \infty\).} 
    \label{equ:rate_mofi}
\end{align}
\end{corollary}

The proof of Corollary \ref{cor:rate_mofi_special} is similar to the proof of Theorem 2 in \cite{guo2024rkhs}, with additional use of the covariance structure of \(\mathscr{A}\) under the special case in \eqref{equ:A_special}, and is therefore omitted. 

\begin{remark}
    \noindent \textbf{(i):} If all true signals are simple, that is, \(q_1=0\), then the second term in \eqref{equ:rate_mofi} vanishes, and the prediction risk reduces to the parametric rate \(O_p(qM_0/n)\), corresponding to a linear regression model with \(qM_0\) predictors.

    \noindent \textbf{(ii):} Under the setting in Remark 4 (iii), where $q = n^{\varsigma}$, $\log p = O\bigl(n^{1-2\varsigma}\bigr)$, $0 < \varsigma < \tfrac14$;
and when the tuning parameters \(\lambda_1,\lambda_2,\bar\lambda_1,\bar\lambda_2\) are chosen at their lower-order bounds, we obtain
\begin{align*}
    \mathcal{R}\!\left(\widehat{\boldsymbol{\beta}}^{(\mathrm{fenet})}\right) = O_p(1), \qquad
\mathcal{R}\!\left(\widehat{\boldsymbol{\beta}}^{(\mathrm{MoFI})}\right) = O_p\!\left(n^{-(\frac12-2\varsigma)}\right) = o_p(1).
\end{align*}
Thus, under this ultra-high-dimensional regime, the Step-Two estimator attains a faster rate for the excess prediction risk than the fully nonparametric functional elastic-net, provided that the signal sets are correctly identified. In particular, when 
$q$ is fixed, this rate simplifies to $O_p(n^{-1/2})$.
\end{remark}

\section{Algorithm}
\label{sec:algorithm}

We assume that the $p$-dimensional functional predictors are observed at $N$ equally spaced points on the interval $[0,1]$. In other words, the observation points are $\left\{\frac{1}{N}, \frac{2}{N}, \dots, \frac{N-1}{N}, 1\right\}$, and if the data are not originally on this grid, they can be rescaled to $[0,1]$. Before performing any analysis, we center both the functional predictors and the responses. Let
$\mathbf{Y}_n = (Y_1, \dots, Y_n)^{\top}$
denote the response vector, and let
$\mathbf{X}_{nj} = \big( X_{ij}(t/N) \big)_{1\le i \le n}^{1\le t \le N}$ be the $n\times N$ matrix corresponding to the discretized $j$-th functional predictor $(j = 1, \dots, p)$.
We denote by $\mathbf{K}_j$ the $N \times N$ matrix obtained by evaluating the reproducing kernel $K_j$ at the observed points, i.e.,
$\mathbf{K}_j = \bigl( K_j(t/N,s/N) \bigr)_{1 \le t,s \le N}.$
To approximate the $\mathbb{L}_2[0,1]$ integral, we use a Riemann sum over $N$ equally spaced grid points. Hence, we define
\begin{align*}
    \widetilde{\mathbf{X}}_{nj} = \frac{1}{N}\mathbf{X}_{nj}\mathbf{K}_j^{1/2} \quad \mbox{and} \quad 
\mathbf{T}_{nj} = \frac{1}{n}\widetilde{\mathbf{X}}_{nj}^\top \widetilde{\mathbf{X}}_{nj}.
\end{align*}
Furthermore, let $\mathbf{B}_{nj}$ denote the $N \times M_j$ matrix containing the first $M_j$ eigenvectors of $\mathbf{T}_{nj}$ ($M_0 < M_j < N$), and define
$\widetilde{\mathbf{B}}_{nj} = \sqrt{N}\, \mathbf{B}_{nj}$.
As a result, $\mathbf{T}_{nj}$ and $\widetilde{\mathbf{B}}_{nj}$ provide a good approximation to the kernel and the first $M_j$ eigenfunctions of $\Tnjj$.
We specify $\Psi_j = (\Tnjj + \theta \mathscr{I})^{1/2}$
and $\Psi_j^{(1)} = (\Tlnjj + \theta \mathscr{I})^{1/2}$
for the rest of the paper.

{\bf Step-One}: 
By Proposition \ref{lemma:solution_form}, the minimizer of \eqref{equ:mini1}, denoted by $\wh f_j$ (for $j=1, \dots, p$), lies in the subspace spanned by $\{\wt{X}_{1j}, \dots, \wt{X}_{nj}\}$ (for each $j=1, \dots, p$). Consequently, the evaluation of $\wh f_j$ on the $N$ grid points, denoted by $\wh{\mathbf{f}}_{nj}$, can be well approximated by
$\wh{\mathbf{f}}_{nj} = \wt{\mathbf{B}}_{nj} \mathbf{c}_j$,
for some $M_j$-dimensional vector $\mathbf{c}_j$.
As such, (\ref{equ:mini1}) can be rewritten as
\vspace{-0.5em}
\begin{align}\label{equ:mini3}
    \frac{1}{2n} \bigg\| \boldsymbol{Y}_n -  \sum_{j=1}^p \boldsymbol{\Gamma}_{nj}  \mathbf{c}_j  \bigg\|_2 ^2 + \lambda_1 \sum_{j=1}^p \| \mathbf{H}_{nj}^{1/2} \mathbf{c}_j \|_2
 + \frac{\lambda_2}{2} \sum_{j=1}^p \|\mathbf{c}_j \|_2^2, 
\end{align}
where $\boldsymbol{\Gamma}_{nj} = N^{-1}  \wt{\mathbf{X}}_{nj} \wt{\mathbf{B}}_{nj}$ and
$\mathbf{H}_{nj} = n^{-1} \boldsymbol{\Gamma}_{nj}^{\top} \boldsymbol{\Gamma}_{nj} + \theta \ \mathbf{I}_{M_j}$.
It is easy to see that $\mathbf{H}_{nj} = N^{-1}\boldsymbol{\Lambda}_{nj} + \theta \ \mathbf{I}_{M_j}$, where $\boldsymbol{\Lambda}_{nj}$ is a $M_j \times M_j$ diagonal matrix with its diagonal values to be the first $M_j$ eigenvalues of $\mathbf{T}_{nj}$.

We reparameterize the coefficient vectors as 
$\mathbf{b}_j = \mathbf{H}_{nj}^{1/2} \mathbf{c}_j,$
and solve the group elastic-net problem (\ref{equ:mini3}) iteratively using a block coordinate-descent algorithm \citep{guo2024rkhs}. For coordinate $j$, we fix $\mathbf{b}_{j'}$ for all $j' \neq j$, and define the residual (after eliminating the contributions of the other coordinates) as 
$\mathbf{Y}_n^{(j)} = \mathbf{Y}_n - \sum_{j' \neq j} \boldsymbol{\Gamma}_{nj'} \mathbf{H}_{nj'}^{-1/2} \mathbf{b}_{j'}.$
We then define
\begin{align*}
&\boldsymbol{\Omega}_{nj} =  \mathbf{I}_{M_j} + (\lambda_2 - \theta) \mathbf{H}_{nj}^{-1}, \quad \boldsymbol{\varrho}_{nj} = \frac{1}{n} \mathbf{H}_{nj}^{-1/2} \boldsymbol{\Gamma}_{nj}^{\top} \mathbf{Y}_n^{(j)}.
\end{align*}
and update $\mathbf{b}_{j}$ by $\wh{\mathbf{b}}_j = \mathbf{0}$ if $\| \boldsymbol{\varrho}_{nj} \|_2 \le \kappa \lambda_1$; when $\| \boldsymbol{\varrho}_{nj} \|_2 > \kappa \lambda_1$, $\wh{\mathbf{b}}_j$ is the solution to the following equation: 
\begin{align}
 \boldsymbol{\Omega}_{nj} \mathbf{b}_j - \boldsymbol{\varrho}_{nj} + \lambda_1 \mathbf{b}_j\| \mathbf{b}_j \|_2^{-1} = \mathbf{0}.
 \label{equ:solve_d}
\end{align}
Here, $\kappa \ge 1$ induces additional sparsity in the approximation errors.
We can solve $\widehat{\mathbf{b}}_j$ by iteratively updating $\mathbf{b}_j \leftarrow \left( \boldsymbol{\Omega}_{nj} + \lambda_1 \| \mathbf{b}_j \|_2^{-1} \mathbf{I}_{M_j} \right)^{-1} \boldsymbol{\varrho}_{nj}$ until convergence. 
When $\wh{\mathbf{b}}_j$ converges for all $j=1,\dots, p$, the functional coefficients can be estimated by $\wh{\mathbf{f}}_{nj} = \wt{\mathbf{B}}_{nj} \mathbf{H}_{nj}^{-1/2} \wh{\mathbf{b}}_j$; and we have
$\widehat{\mathcal{S}} = \left\{ j \in \{1,\dots,p\} : \wh{\mathbf{b}}_j \neq 0 \right\}$.
For full details, see Algorithm \ref{alg:step-one}.

\begin{algorithm}[ht]
\caption{MoFI-FLR (Step-One)}
\label{alg:step-one}
\begin{algorithmic}[1]
  \STATE Discretize functional predictors $\mathbf{X}_{nj} = \big( X_{ij}(t/N) \big)_{1\le i \le n}^{1\le t \le N}$ and reproducing kernel $\mathbf{K}_j = \bigl( K_j(t/N,s/N) \bigr)_{1 \le t,s \le N}$ on $[0,1]$ for $j=1, \dots, p$.
  \STATE Center responses $\mathbf{Y}_n$ and predictors $\{ \mathbf{X}_{nj} \}_{j=1}^p$.
  \STATE Compute $\widetilde{\mathbf{X}}_{nj}$, $\wt{\mathbf{B}}_{nj}$, $\mathbf{\Gamma}_{nj}$, 
  $\mathbf{H}_{nj}$, and $\boldsymbol{\Omega}_{nj}$ for $j=1, \dots, p$.
  \STATE Initialize $\mathbf{b}_j, \mathbf{b}_j^{\text{new}} = \mathbf{0}$ for $j=1, \dots, p$ and let $\boldsymbol{\eta}_n \leftarrow \mathbf{Y}_n$.
  \WHILE{$ \max_j \| \mathbf{b}_j^{\text{new}} - \mathbf{b}_j \|_2 > \text{tolerance}$}
  \STATE Update $\mathbf{b}_j \leftarrow \mathbf{b}_j^{\text{new}}$ for $j=1, \dots, p$. 
  \FOR{$j = 1, \dots, p$}
      \IF{$j=1$}
      \STATE Update $\boldsymbol{\eta}_n \leftarrow \boldsymbol{\eta}_n - \mathbf{\Gamma}_{np} \mathbf{H}_{np}^{-1/2} \mathbf{b}_p^{\text{new}} + \mathbf{\Gamma}_{n1} \mathbf{H}_{n1}^{-1/2} \mathbf{b}_1$.
    \ELSE
      \STATE Update $\boldsymbol{\eta}_n \leftarrow \boldsymbol{\eta}_n - \mathbf{\Gamma}_{n,j-1} \mathbf{H}_{n,j-1}^{-1/2} \mathbf{b}_{j-1}^{\text{new}} + \mathbf{\Gamma}_{nj} \mathbf{H}_{nj}^{-1/2} \mathbf{b}_j$.
    \ENDIF
    \STATE Compute $\boldsymbol{\varrho}_{nj} = n^{-1} \mathbf{H}_{nj}^{-1/2} \boldsymbol{\Gamma}_{nj}^{\top} \boldsymbol{\eta}_n$.
      \IF{$\| \boldsymbol{\varrho}_{nj} \|_2 \le \kappa \lambda_1$}
      \STATE Update $\mathbf{b}_j^{\text{new}} \leftarrow 0$
    \ELSE
      \STATE Update $\mathbf{b}_j^{\text{new}} \leftarrow \left( \boldsymbol{\Omega}_{nj} + \lambda_1 \| \mathbf{b}_j \|_2^{-1} \mathbf{I}_{M_j} \right)^{-1} \boldsymbol{\varrho}_{nj}$ until convergence.
    \ENDIF
  \ENDFOR
  \ENDWHILE
  \RETURN $\wh{\mathbf{f}}_{nj} = \wt{\mathbf{B}}_{nj} \mathbf{H}_{nj}^{-1/2} \mathbf{b}_j^{\text{new}}$ and $\widehat{\mathcal{S}} = \left\{ j \in \{1,\dots,p\} : \mathbf{b}_j^{\text{new}} \neq 0 \right\}$.
\end{algorithmic}
\end{algorithm}

{\bf Step-Two}: 
We denote by $\mathbf{K}_{0j}$ and $\mathbf{K}_{1j}$ the $N \times N$ matrices obtained by evaluating the reproducing kernels $K_{0j}$ and $K_{1j}$ at the observed points. 
Let $\boldsymbol{\Phi}_{j}$ be the $N \times M_0$ matrix whose columns consist of the $M_0$ eigenvectors of $\mathbf{K}_{0j}$. Define 
\begin{align*}
    &\widetilde{\boldsymbol{\Phi}}_{j} = \sqrt{N}\, \boldsymbol{\Phi}_{j}, \quad 
    \wt{\mathbf{Z}}_{j} = \frac{1}{N}\mathbf{X}_{nj} \widetilde{\boldsymbol{\Phi}}_{j}, \quad
    \widetilde{\mathbf{X}}_{nj}^{(1)} = \frac{1}{N}\mathbf{X}_{nj}\mathbf{K}_{1j}^{1/2}, \quad \mbox{and} \quad 
\mathbf{T}_{1nj} = \frac{1}{n} \left(\widetilde{\mathbf{X}}_{nj}^{(1)} \right)^\top \widetilde{\mathbf{X}}_{nj}^{(1)}.
\end{align*}
Furthermore, let $\mathbf{B}_{1nj}$ denote the $N \times (M_j - M_0)$ matrix containing the first $M_j - M_0$ eigenvectors of $\mathbf{T}_{1nj}$, and define
$\widetilde{\mathbf{B}}_{1nj} = \sqrt{N}\, \mathbf{B}_{1nj}$.
As a result, $\mathbf{T}_{1nj}$ and $\widetilde{\mathbf{B}}_{1nj}$ provide a good approximation to the kernel and eigenfunctions of $\Tlnjj$.

By Proposition \ref{lemma:solution_form}, the evaluation of $\wh f_j^{(1)}$ on the $N$ grid points, denoted by $\wh{\mathbf{f}}_{nj}^{(1)}$, can be well approximated by
$\wh{\mathbf{f}}_{nj}^{(1)} = \wt{\mathbf{B}}_{1nj} \mathbf{c}_j^{(1)}$,
for some $(M_j-M_0)$-dimensional vector $\mathbf{c}_j^{(1)}$.
As a result, (\ref{equ:mini2}) can be rewritten as
\vspace{-0.5em}
\begin{align}\label{equ:mini4}
    \frac{1}{2n} \bigg\| \boldsymbol{Y}_n - \sum_{j \in \wh{\cal S}} \wt{\mathbf{Z}}_{j} \mathbf{a}_j - \sum_{ j \in \wh{\cal S} } \boldsymbol{\Gamma}_{1nj}  \mathbf{c}_j^{(1)}  \bigg\|_2 ^2 + \widebar{\lambda}_1 \sum_{ j \in \wh{\cal S} } \left\| \mathbf{H}_{1nj}^{1/2} \mathbf{c}_j^{(1)} \right\|_2
 + \frac{\widebar{\lambda}_2}{2} \sum_{ j \in \wh{\cal S} } \left\|\mathbf{c}_j^{(1)} \right\|_2^2, 
\end{align}
where $\boldsymbol{\Gamma}_{1nj} = N^{-1}  \wt{\mathbf{X}}_{nj}^{(1)} \wt{\mathbf{B}}_{1nj}$ and
$\mathbf{H}_{1nj} = n^{-1} \boldsymbol{\Gamma}_{1nj}^{\top} \boldsymbol{\Gamma}_{1nj} + \theta \ \mathbf{I}_{M_j-M_0}$. Similar to $\mathbf{H}_{nj}$, $\mathbf{H}_{1nj}$ is also a diagonal matrix.

We reparameterize the coefficient vectors as
$\mathbf{b}_j^{(1)} = \mathbf{H}_{1nj}^{1/2} \mathbf{c}_j^{(1)},$
and subsequently solve the group elastic-net problem (\ref{equ:mini4}) iteratively using a block coordinate descent algorithm analogous to that employed in Step-One.
Define $\wt{\mathbf{Z}}_{\wh{\cal S}} = (\wt{\mathbf{Z}}_{j})_{j \in \wh{\cal S}}$.
We first fix $\mathbf{b}_j^{(1)}$ ($j \in \wh{\cal S}$)
and update $\mathbf{a}_{\wh{\cal S}} = (\mathbf{a}_{j}^{\top})^{\top}_{j \in \wh{\cal S}}$ by
\begin{align*}
    \wh{\mathbf{a}}_{\wh{\cal S}} = \left(  \wt{\mathbf{Z}}_{\wh{\cal S}}^{\top}  \wt{\mathbf{Z}}_{\wh{\cal S}} \right)^{-1} \wt{\mathbf{Z}}_{\wh{\cal S}}^{\top} \bigg(\boldsymbol{Y}_n - \sum_{ j \in \wh{\cal S} } \boldsymbol{\Gamma}_{1nj} \mathbf{H}_{1nj}^{-1/2} \wh{\mathbf{b}}_j^{(1)}  \bigg).
\end{align*}
We need $M_0 \cdot |\wh{\cal S}| \le n$ in order to obtain a unique solution. 

Next, for coordinate $j \in \wh{\cal S}$, we fix $\wh{\mathbf{a}}_{\wh{\cal S}}$ and $\mathbf{b}_{j'}^{(1)}$ for all $j' \in \wh{\cal S} / \{j\}$, and define the residual (after eliminating the contributions of the other coordinates) as 
$\wt{\mathbf{Y}}_n^{(j)} = \mathbf{Y}_n - \wt{\mathbf{Z}}_{\wh{\cal S}} \wh{\mathbf{a}}_{\wh{\cal S}} - \sum_{j' \in \wh{\cal S} / \{j\} } \boldsymbol{\Gamma}_{1nj'} \mathbf{H}_{1nj'}^{-1/2} \mathbf{b}_{j'}^{(1)}.$
We then define
\begin{align*}
&\boldsymbol{\Omega}_{1nj} =  \mathbf{I}_{M_j-M_0} + (\bar\lambda_2 - \theta) \mathbf{H}_{1nj}^{-1}, \quad \boldsymbol{\varrho}_{1nj} = \frac{1}{n} \mathbf{H}_{1nj}^{-1/2} \boldsymbol{\Gamma}_{1nj}^{\top} \wt{\mathbf{Y}}_n^{(j)}.
\end{align*}
and update $\mathbf{b}_{j}^{(1)}$ by $\wh{\mathbf{b}}_j^{(1)} = \mathbf{0}$ if $\| \boldsymbol{\varrho}_{1nj} \|_2 \le \widebar{\kappa} \widebar{\lambda}_1$; when $\| \boldsymbol{\varrho}_{1nj} \|_2 > \widebar{\kappa} \widebar{\lambda}_1$, $\wh{\mathbf{b}}_j^{(1)}$ is the solution to the following equation: 
\begin{align}
 \boldsymbol{\Omega}_{1nj} \mathbf{b}_j^{(1)} - \boldsymbol{\varrho}_{1nj} + \widebar{\lambda}_1 \mathbf{b}_j^{(1)}\| \mathbf{b}_j^{(1)} \|_2^{-1} = \mathbf{0}.
 \label{equ:solve_d1}
\end{align}
When $\wh{\mathbf{a}}_{\wh{\cal S}}$ and all $\wh{\mathbf{b}}_j^{(1)}$ $(j \in \wh{\cal S})$ converge, we have $\widehat{\mathcal{S}}_0 = \left\{ j \in \wh{\calS} : \ \wh{\mathbf{b}}_j^{(1)} = 0 \right\}$.
To that end, the refined functional coefficients evaluated on the $N$ grids can be estimated by 
\begin{align}
    \wh{\beta}_{j}\left(\frac{1}{N}, \frac{2}{N}, \dots, \frac{N-1}{N}, 1 \right) = \wt{\boldsymbol{\Phi}}_{j} \wh{\mathbf{a}}_j + \frac{1}{N} \mathbf{K}_{1j} \wt{\mathbf{B}}_{1nj} \mathbf{H}_{1nj}^{-1/2} \wh{\mathbf{b}}_j^{(1)}.
\end{align}
For more details, see Algorithm \ref{alg:step-two}.

\begin{algorithm}[ht]
\caption{MoFI-FLR (Step-Two)}
\label{alg:step-two}
\begin{algorithmic}[1]
  \STATE Re-label the elements in $\wh{\calS}$ as $\{1, \dots, \wh{q}\}$, evaluating $\mathbf{K}_{0j}$ and $\mathbf{K}_{1j}$ at the $N \times N$ grids. 
  \STATE Compute $\wt{\boldsymbol{\Phi}}_{j}$, $\wt{\mathbf{Z}}_{\wh{\calS}}$, $\widetilde{\mathbf{X}}_{nj}^{(1)}$, $\wt{\mathbf{B}}_{1nj}$, $\mathbf{\Gamma}_{1nj}$, 
  $\mathbf{H}_{1nj}$, and $\boldsymbol{\Omega}_{1nj}$ for $j=1, \dots, \wh{q}$.
  \STATE Initialize $\mathbf{a}_{\wh{\calS}}, \mathbf{a}_{\wh{\calS}}^{\text{new}} = \mathbf{0}$, $\mathbf{b}_j^{(1)}, \mathbf{b}_j^{(1)\text{new}} = \mathbf{0}$,  for $j=1, \dots, \wh{q}$, and let $\boldsymbol{\eta}_n \leftarrow \mathbf{Y}_n$.
  \WHILE{$\max\left( \| \mathbf{a}_{\wh{\calS}}^{\text{new}} - \mathbf{a}_{\wh{\calS}} \|_2, \ \max_j \| \mathbf{b}_j^{(1)\text{new}} - \mathbf{b}_j^{(1)} \|_2 \right) > \text{tolerance}$ }
  \STATE Update $\mathbf{a}_{\wh{\calS}} \leftarrow \mathbf{a}_{\wh{\calS}}^{\text{new}}$, $\mathbf{a}_{\wh{\cal S}}^{\text{new}} \leftarrow \left(  \wt{\mathbf{Z}}_{\wh{\cal S}}^{\top}  \wt{\mathbf{Z}}_{\wh{\cal S}} \right)^{-1} \wt{\mathbf{Z}}_{\wh{\cal S}}^{\top} \mathbf{r}_n$, and $\mathbf{b}_j^{(1)} \leftarrow \mathbf{b}_j^{(1)\text{new}}$ for $j=1, \dots, \wh{q}$.
  \FOR{$j = 1, \dots, \wh{q}$}
      \IF{$j=1$}
      \STATE Update $\boldsymbol{\eta}_n \leftarrow \boldsymbol{\eta}_n - \wt{\mathbf{Z}}_{\wh{\cal S}} \mathbf{a}_{\wh{\calS}}^{\text{new}} + \mathbf{\Gamma}_{1n1} \mathbf{H}_{1n1}^{-1/2} \mathbf{b}_1$.
    \ELSE
      \STATE Update $\boldsymbol{\eta}_n \leftarrow \boldsymbol{\eta}_n - \mathbf{\Gamma}_{1n,j-1} \mathbf{H}_{1n,j-1}^{-1/2} \mathbf{b}_{j-1}^{(1)\text{new}} + \mathbf{\Gamma}_{1nj} \mathbf{H}_{1nj}^{-1/2} \mathbf{b}_j^{(1)}$.
    \ENDIF
    \STATE Compute $\boldsymbol{\varrho}_{1nj} = n^{-1} \mathbf{H}_{1nj}^{-1/2} \boldsymbol{\Gamma}_{1nj}^{\top} \boldsymbol{\eta}_n$.
      \IF{$\| \boldsymbol{\varrho}_{1nj} \|_2 \le \bar\kappa \bar\lambda_1$}
      \STATE Update $\mathbf{b}_j^{(1)\text{new}} \leftarrow 0$
    \ELSE
      \STATE Update $\mathbf{b}_j^{(1)\text{new}} \leftarrow \left( \boldsymbol{\Omega}_{1nj} + \bar\lambda_1 \| \mathbf{b}_j^{(1)} \|_2^{-1} \mathbf{I}_{M_j-M_0} \right)^{-1} \boldsymbol{\varrho}_{1nj}$ until convergence.
    \ENDIF
  \ENDFOR
  \ENDWHILE
  \RETURN $\wh{\boldsymbol{\beta}}_{nj} = \wt{\boldsymbol{\Phi}}_{j} \mathbf{a}_j^{\text{new}} + N^{-1} \mathbf{K}_{1j} \wt{\mathbf{B}}_{1nj} \mathbf{H}_{1nj}^{-1/2} \mathbf{b}_j^{(1)\text{new}}$, $\widehat{\mathcal{S}}_0 = \{ j \in \wh{\calS} : \mathbf{b}_j^{(1)\text{new}} = \mathbf{0} \}$.
\end{algorithmic}
\end{algorithm}

\begin{remark}
The computational burden of our method is significantly reduced by exploiting the problem's structure. 
First, each iteration of Algorithms \ref{alg:step-one} and \ref{alg:step-two} is highly efficient, 
as the diagonal structure eliminates the need for matrix inversion. 
Second, the computational complexity for calculating the partial residuals $\mathbf{Y}_n^{(j)}$ and
$\wt{\mathbf{Y}}_n^{(j)}$ in the two algorithms can be greatly reduced by observing that most components of the partial residuals remain unchanged during the coordinate descent updates between consecutive coordinates; see the algorithm descriptions for details on the update of $\boldsymbol{\eta}_n$ (a unified representation of ${\mathbf{Y}}_n^{(j)}$ and $\wt{\mathbf{Y}}_n^{(j)}$).
\end{remark}

\noindent \textbf{Tuning parameter selection:}
    We select the tuning parameters via cross-validation over a discrete grid. Similar approaches have been used by \cite{zhang2011linear} for structure discovery in partially linear models and by \cite{XueYao2021} for high-dimensional functional linear regressions. 
    A useful strategy is to begin with a relatively coarse grid and then refine the search over a narrower range if needed. In particular, a finer search may be used for the more sensitive tuning parameters, whereas relatively coarse grids are often adequate for the less sensitive ones.

    Recall that the covariance operator of \(\widetilde X_{ij}\) is approximated using its first \(M_j\) eigenfunctions. As noted by \citet{guo2024rkhs}, both variable selection and prediction performance are generally not sensitive to the choice of \(M_j\), even when \(M_j\) is chosen to be fairly large.
    In particular, eigenfunctions associated with small eigenvalues tend to have only a negligible effect on variable selection and prediction. Accordingly, \(M_j\) may be fixed in advance or selected over a relatively coarse grid.

    We reparameterize $\lambda_1=\alpha\lambda$, $\lambda_2=(1-\alpha)\lambda$, $\bar\lambda_1=\bar\alpha\bar\lambda$, $\bar\lambda_2=(1-\bar\alpha)\bar\lambda$.
    In Step-One, variable selection is relatively sensitive to the choice of \((\lambda,\alpha)\), so we tune \(\lambda\) and \(\alpha\) over relatively fine grids. In Step-Two, as noted in Remark~4(ii), \(\bar\lambda\) is typically smaller than \(\lambda\). Therefore, we restrict the search for \(\bar{\lambda}\) to values no larger than the optimal \(\lambda\) selected in Step-One.
    Similarly, we search for \(\bar\alpha\) over a range centered around the optimal value of \(\alpha\).

    For selecting \(\theta\), which appears in \(\Psi_j\) and \(\Psi_j^{(1)}\), recall that $\Psi_j = \bigl(\mathscr T_n^{(j,j)} + \theta \mathscr I \bigr)^{1/2}$.
    It is important that \(\theta\) should not dominate \(\mathscr T_n^{(j,j)}\), since the sample covariance operator plays an essential role in maintaining good prediction performance \citep{XueYao2021}. Therefore, in Step-One, we restrict the search for \(\theta\) to values smaller than the largest eigenvalue among \(\{\mathscr T_n^{(j,j)}\}_{j=1}^p\), and then keep it fixed in Step-Two.

\section{Simulation Studies}\label{sec:simulations}

Consider the cosine basis $\phi_1(t) = 1$, $\phi_k(t) = \sqrt{2} \cos ((k-1)\pi t)$, $k > 1$.
We simulate the functional predictors according to
\begin{align*}
X_{ij}(t) = \sum_{k=1}^{30} z_{ijk}\nu_k^{1/2}\,\phi_k(t), \quad i = 1,\ldots,n,\quad j = 1,\ldots,p,
\end{align*}
where for each $k$ the vector $\boldsymbol{z}_{i \cdot k} =(z_{i1k}, \ldots, z_{ipk})^\top$ independently and identically follows an multivariate normal distribution $N(\boldsymbol{0}, \boldsymbol{\Sigma}_p)$, and $\boldsymbol{\Sigma}_p$ is an autoregressive correlation matrix with the $(j,k)$th entry being $\rho^{|j-k|}$ for $1 \leq k, j \leq p$. We fix $\rho = 0.5$ in the entire study. The coefficient $\nu_k$ follows $\nu_k= \exp(-k/4)$ for $k \ge 1$.
The response variable $Y_i$ is generated by the high-dimensional functional linear regression model \eqref{e:model}. In this model, the error terms satisfy $\epsilon_i \sim N(0, \sigma^2)$.
We consider two scenarios for the coefficient functions.

\textbf{Scenario I.} This scenario aims to identify constant coefficient functions. Let the true coefficient functions be
\begin{equation}
  \beta_{0j}(t) = 4 \sum_{k=1}^{30} (-1)^{u_{jk}} \gamma_{jk} \;\nu_k\;  \phi_k(t), \label{equ:beta_simu}
\end{equation}
where $u_{jk}\sim\mathrm{Bernoulli}(0.5)$ are independent.  
We assume exactly the first $q$ coefficients are nonzero, and that among these, the first $q_0 = \big\lfloor q/2\big\rfloor$
are constant functions.  Equivalently, $\gamma_{jk}=I(k=1)$ for
$j=1, \dots, q_0$ and $\gamma_{jk}=1$ for $j = q_0+1, \dots, q$.
Suppose $K_{0j} = K_0^{\rm I}$ and $K_{1j} = K_1^{\rm I}$, and denote $K = K_0^{\rm I} + K_1^{\rm I}$. Suppose the RKHS $\mathbb{H}(K)$ is endowed with the norm 
\begin{align}
    \| \beta \|_{\mathbb{H}(K)}^2 =  \pi^4 \left(\int \beta \right)^2 + \int (\beta^{''})^2. \label{equ:beta_rkhs_norm}
\end{align}
Then, we have $\| \beta \|_{\mathbb{H}(K)}^2 = \pi^4 \sum_{k\ge 1} (k^4 + 1) f_{k}^2$ for any $\beta = \sum_{k \ge 1} f_k \phi_k \in \mathbb{H}(K)$.
We can show that $K_0^{\rm I}(s,t) = 1/\pi^4$ and 
\begin{align*}
    K_1^{\rm I}(s,t)  =-\frac{1}{3}\Big\{ B_{4}(|s-t| / 2)+B_{4}\{(s+t) / 2\} \Big\},
\end{align*}
where $B_k(\cdot)$ is the $k$-th Bernoulli polynomial, and hence $B_4(t) = t^4 - 2t^3 + t^2 - 1/30$.

\textbf{Scenario II.} In this scenario, we aim to identify periodic coefficient functions with a certain frequency. To this end, we retain the form of $\beta_j$ from \eqref{equ:beta_simu} but set
$\gamma_{jk}=I(k=2)$ for
$(j=1, \dots, q_0)$,
thereby restricting each coefficient to oscillate with a fundamental period to be $2$ (i.e.\ low‐frequency variation).
Under the RKHS norm defined in \eqref{equ:beta_rkhs_norm},
\begin{align*}
    K_0^{\rm II}(s,t) = 2 \pi^{-4} \cos(\pi s) \cos(\pi t). 
\end{align*}
it follows that $K_1^{\rm II} = K - K_0^{\rm II}$.

\textbf{Scenario III.} The simple coefficient functions in Scenario III are linear combinations of those in Scenario I and II. Specifically, we retain the form of $\beta_j$ from \eqref{equ:beta_simu} and set
$\gamma_{jk}=I(k \in \{1, 2\})$ for
$(j=1, \dots, q_0)$. 
Under the RKHS norm defined in \eqref{equ:beta_rkhs_norm}, we have $K_0^{\rm III} = K_0^{\rm I} + K_0^{\rm II}$ and $K_1^{\rm III} = K - K_0^{\rm III}$.

We adopt the configuration $(p,q,q_0)=(100, 10, 5)$ throughout the study and perform 200 simulation replications. In each replication, we generate a training sample of size $n \in \{125,250\}$ and add Gaussian noise with standard deviation $\sigma \in \{1,2,4,8\}$ to control the signal-to-noise ratio. An independent test set of size $n_{\mathrm{test}}=1000$ is also simulated to assess predictive performance.
We apply MoFI-FLR to the simulated data sets and make a comparison with the functional elastic-net (fENet) and the refined estimator of functional elastic-net
\citep[fENet-refine,][]{guo2024rkhs}.
For both fENet and fENet‑refine, tuning parameters are selected via a grid search by five‐fold cross‐validation, minimizing the average mean squared prediction error across folds in the training set.  For MoFI, we similarly employ five‐fold cross‐validation to choose the Step‑One tuning parameters.  In Step-Two, we explore two strategies: (a) Select the second‑step penalty by five‑fold cross‑validation (MoFI-optim); (b) Reuse the stage‑one penalty as the second‑stage penalty (MoFI-fix).
We assess predictive performance via the relative excess risk (RER), defined as
\begin{align*}
    { \E^* \{ \sum_{j=1}^p  \langle X_j^\ast, (\hat \beta_j- \beta_{j0}) \rangle\}^2  / \E^* \{ \sum_{j=1}^p  \langle X_j^\ast, \beta_{j0} \rangle\}^2 },
\end{align*}
where $\E^*$ denotes the empirical expectation over the testing set.
We report the mean and $(5\%, 95\%)$ quantile of the RER across the 200 simulation replicates in Table \ref{tab:rer}.

Across all three scenarios, the RER increases markedly as the noise level $\sigma$ grows from $1$ to $8$ and decreases when the sample size doubles from $n = 125$ to $n = 250$, reflecting the anticipated impact of signal‐to‐noise ratio (SNR) and sample size on prediction accuracy. 
In low-noise settings ($\sigma=1, 2$), MoFI-optim yields the lowest mean RER compared to fENet and fENet-refine for both sample sizes. This superiority persists at moderate noise ($\sigma=4$), especially when $n=250$. Under high noise ($\sigma=8$), however, both MoFI-fix and MoFI-optim incur slightly higher RER compared to the fENet or fENet-refine. This finding is consistent with the simulations of \cite{zhang2011linear}.
In Scenarios I and II, both MoFI-fix and MoFI-optim exhibit similar predictive performance, whereas in Scenario III, MoFI-fix is the worst among the four methods. This is because the $\mathbb{H}(K_1^{\rm III})$ only contains the higher order cosine basis, requiring a higher $\ell_2$-penalty ratio to ensure sufficient smoothness.
Overall, MoFI-optim delivers the best prediction performance in low-to-moderate noise regimes and remains competitive even when noise is high and sample size is small.

\begin{table}[ht]
\centering
\caption{Mean and $(5\%, 95\%)$ quantile of the relative excess risk (RER) computed over 200 simulation replicates under three scenarios.}
\resizebox{\textwidth}{!}{
\label{tab:rer}
\begin{tabular}{@{} cccccc c @{}}
\toprule
Scenario & \(n\) & \(\sigma\) & fENet & fENet-refine & MoFI-fix & MoFI-optim \\
\midrule
\multirow{8}{*}{I}
  & \multirow{4}{*}{125} & 1 & $0.0194 \ (0.0111, 0.0271)$ & $0.0176 \ (0.0106, 0.0229)$ & $0.0115 \ (0.0054, 0.0144)$ & $ \mathbf{0.0112} \ (0.0053, 0.0134)$ \\
  &                      & 2 & $0.0536 \ (0.0279, 0.1199)$ & $0.0519 \ (0.0239, 0.1374)$ & $0.0355 \ (0.0146, 0.0918)$ & $\mathbf{0.0349} \ (0.0144, 0.1230)$ \\
  &                      & 4 & $0.1895 \ (0.1031, 0.3297)$ & $0.1957 \ (0.1086, 0.3373)$ & $0.1858 \ (0.0718, 0.3418)$ & $\mathbf{0.1740} \ (0.0675, 0.3284)$ \\
  &                      & 8 & $0.5413 \ (0.3194, 0.8469)$ & $\mathbf{0.5229} \ (0.3126, 0.7761)$ & $0.5806 \ (0.3580, 0.9797)$ & $0.6049 \ (0.3210, 0.9853)$ \\ [1ex]
  & \multirow{4}{*}{250} & 1 & $0.0064 \ (0.0049, 0.0086)$ & $0.0056 \ (0.0042, 0.0073)$ & $0.0036 \ (0.0026, 0.0049)$ & $\mathbf{0.0033} \ (0.0023, 0.0047)$ \\
  &                      & 2 & $0.0178 \ (0.0126, 0.0239)$ & $0.0176 \ (0.0123, 0.0241)$ & $0.0114 \ (0.0075, 0.0158)$ & $\mathbf{0.0112} \ (0.0077, 0.0159)$ \\
  &                      & 4 & $0.0501 \ (0.0317, 0.0830)$ & $0.0537 \ (0.0345, 0.0849)$ & $0.0385 \ (0.0213, 0.0863)$ & $\mathbf{0.0363} \ (0.0213, 0.0678)$ \\
  &                      & 8 & $\mathbf{0.2471} \ (0.1531, 0.3674)$ & $0.2639 \ (0.1629, 0.3929)$ & $0.2799 \ (0.1615, 0.4270)$ & $0.2551 \ (0.1489, 0.4019)$ \\ [1ex]

\multirow{8}{*}{II}
  & \multirow{4}{*}{125} & 1 & $0.0190 \ (0.0093, 0.0606)$ & $0.0197 \ (0.0097, 0.0599)$ & $0.0120 \ (0.0045, 0.0433)$ & $\mathbf{0.0116} \ (0.0047, 0.0435)$ \\
  &                      & 2 & $0.0507 \ (0.0232, 0.0898)$ & $0.0533 \ (0.0246, 0.0906)$ & $\mathbf{0.0395} \ (0.0140, 0.0785)$ & $0.0398 \ (0.0137, 0.0820)$ \\
  &                      & 4 & $0.1430 \ (0.0893, 0.2440)$ & $0.1523 \ (0.0964, 0.2712)$ & $0.1417 \ (0.0624, 0.2558)$ & $\mathbf{0.1401} \ (0.0768, 0.2567)$ \\
  &                      & 8 & $0.4386 \ (0.2197, 0.7432)$ & $\mathbf{0.4202} \ (0.2274, 0.6710)$ & $0.4934 \ (0.2462, 0.8986)$ & $0.4880 \ (0.2459, 0.8932)$ \\ [1ex]
  & \multirow{4}{*}{250} & 1 & $0.0054 \ (0.0040, 0.0073)$ & $0.0050 \ (0.0037, 0.0065)$ & $0.0032 \ (0.0023, 0.0043)$ & $\mathbf{0.0029} \ (0.0019, 0.0041)$ \\
  &                      & 2 & $0.0152 \ (0.0106, 0.0209)$ & $0.0157 \ (0.0110, 0.0211)$ & $\mathbf{0.0103} \ (0.0073, 0.0142)$ & $\mathbf{0.0103} \ (0.0070, 0.0155)$ \\
  &                      & 4 & $0.0541 \ (0.0282, 0.0893)$ & $0.0586 \ (0.0320, 0.0901)$ & $0.0449 \ (0.0194, 0.0832)$ & $\mathbf{0.0448} \ (0.0198, 0.0846)$ \\
  &                      & 8 & $\mathbf{0.1803} \ (0.1089, 0.2969)$ & $0.1920 \ (0.1109, 0.3193)$ & $0.2067 \ (0.1058, 0.3478)$ & $0.1840 \ (0.1023, 0.3261)$ \\ [1ex]
  \multirow{8}{*}{III}
  & \multirow{4}{*}{125} & 1 & $0.0154 \ (0.0088, 0.0225)$ & $0.0129 \ (0.0071, 0.0171)$ & $0.0173 \ (0.0044, 0.0557)$ & $\mathbf{0.0086} \ (0.0046, 0.0127)$ \\
  &                      & 2 & $0.0369 \ (0.0226, 0.0637)$ & $0.0334 \ (0.0194, 0.0530)$ & $0.0617 \ (0.0282, 0.1067)$ & $\mathbf{0.0284} \ (0.0147, 0.0475)$ \\
  &                      & 4 & $0.1205 \ (0.0573, 0.2207)$ & $0.1179 \ (0.0528, 0.2091)$ & $0.1456 \ (0.0803, 0.2473)$ & $\mathbf{0.1158} \ (0.0483, 0.2216)$ \\
  &                      & 8 & $\mathbf{0.3671} \ (0.2133, 0.6183)$ & $0.3729 \ (0.2001, 0.6345)$ & $0.4536 \ (0.2393, 0.7689)$ & $0.4412 \ (0.2407, 0.7381)$ \\[1ex]
  & \multirow{4}{*}{250} & 1 & $0.0046 \ (0.0034, 0.0061)$ & $0.0040 \ (0.0030, 0.0054)$ & $0.0028 \ (0.0021, 0.0039)$ & $\mathbf{0.0027} \ (0.0020, 0.0039)$ \\
  &                      & 2 & $0.0138 \ (0.0101, 0.0191)$ & $0.0138 \ (0.0102, 0.0192)$ & $0.0188 \ (0.0066, 0.0446)$ & $\mathbf{0.0086} \ (0.0060, 0.0132)$ \\
  &                      & 4 & $0.0378 \ (0.0256, 0.0566)$ & $0.0391 \ (0.0274, 0.0564)$ & $0.0598 \ (0.0284, 0.1013)$ & $\mathbf{0.0329} \ (0.0212, 0.0501)$ \\
  &                      & 8 & $\mathbf{0.1518} \ (0.0812, 0.2639)$ & $0.1572 \ (0.0801, 0.2900)$ & $0.1736 \ (0.1032, 0.2846)$ & $0.1651 \ (0.0855, 0.3093)$ \\
\bottomrule
\end{tabular}
}
\end{table}

For the proposed MoFI, we use false positive rate (FPR) and false negative rate (FNR), defined as FPR$=|\wh \calS \cap \calS^c| / |\calS^c| $ and FNR$=|\wh \calS^c \cap \calS| / |\calS| $, to assess the variable selection performance. We use the rate of simple effects incorrectly identified as complex effects (denoted as $r(0\!\to\!1)= |\wh \calS_1 \cap \calS_0| / |\calS_0|$), and the rate of complex effects incorrectly identified as simple effects (denoted as $r(1\!\to\!0)= |\wh \calS_0 \cap \calS_1| / |\calS_1|$) to assess the form identification performance. The results can be found in Table \ref{tab:vs_fi}.

\begin{table}[ht]
\footnotesize
\centering
\caption{Variable‐selection and form‐identification metrics over 200 simulation replicates under three scenarios.}
\label{tab:vs_fi}
\begin{tabular}{@{} c c c c c c c c c @{}}
\toprule
\multirow{2}{*}{Scenario} & \multirow{2}{*}{$n$} & \multirow{2}{*}{$\sigma$} & \multirow{2}{*}{FPR$\times 100\%$} & \multirow{2}{*}{FNR$\times 100\%$} 
  & \multicolumn{2}{c}{$r(0\!\to\!1)\times 100\%$} & \multicolumn{2}{c}{$r(1\!\to\!0)\times 100\%$} \\
\cmidrule(lr){6-7} \cmidrule(lr){8-9}
 & & & & & fix & optim & fix & optim \\
 \midrule
\multirow{8}{*}{I}
  & \multirow{4}{*}{125}
    & 1 & 0.19 & 0.20 & 0.10 & 9.00 & 0.00 & 0.00 \\
  & 
    & 2 & 0.81 & 1.40 & 2.50 & 9.70 & 0.80 & 0.10 \\
  & 
    & 4 & 2.67 & 13.30 & 8.80 & 30.90 & 15.50 & 2.20 \\
  & 
    & 8 & 4.55 & 39.58 & 14.79 & 29.94 & 29.09 & 7.64 \\ [1.5ex]
  & \multirow{4}{*}{250}
    & 1 & 0.00 & 0.00 & 0.00 & 2.60 & 0.00 & 0.00 \\
  & 
    & 2 & 0.18 & 0.00 & 0.30 & 3.70 & 0.00 & 0.00 \\
  & 
    & 4 & 0.18 & 0.50 & 0.10 & 6.00 & 1.40 & 0.00 \\
  & 
    & 8 & 1.16 & 27.80 & 1.50 & 25.10 & 30.20 & 3.20 \\ [1.5ex]
\multirow{8}{*}{II}
  & \multirow{4}{*}{125}
    & 1 & 0.41 & 1.55 & 0.80 & 8.20 & 0.00 & 0.00 \\
  & 
    & 2 & 0.60 & 8.85 & 2.30 & 9.40 & 0.10 & 0.00 \\
  & 
    & 4 & 1.74 & 27.30 & 5.90 & 22.20 & 4.50 & 0.40 \\
  & 
    & 8 & 4.96 & 45.62 & 13.37 & 23.03 & 13.37 & 2.25 \\ [1.5ex]
  & \multirow{4}{*}{250}
    & 1 & 0.01 & 0.00 & 0.00 & 2.50 & 0.00 & 0.00 \\
  & 
    & 2 & 0.18 & 0.05 & 0.00 & 6.20 & 0.00 & 0.00 \\
  & 
    & 4 & 0.24 & 9.20 & 0.20 & 5.40 & 0.20 & 0.00 \\
  & 
    & 8 & 0.69 & 41.15 & 0.50 & 10.70 & 12.70 & 0.20 \\ [1.5ex]
    \multirow{8}{*}{III}
  & \multirow{4}{*}{125}
    & 1 & 0.16 & 0.10 & 3.60 & 19.90 & 8.40 & 0.00 \\
  & 
    & 2 & 0.30 & 0.25 & 4.90 & 40.20 & 44.10 & 0.90 \\
  & 
    & 4 & 1.19 & 3.75 & 4.70 & 51.10 & 75.10 & 14.70 \\
  & 
    & 8 & 4.50 & 21.50 & 5.30 & 20.40 & 66.80 & 51.40 \\[1.5ex]
  & \multirow{4}{*}{250}
    & 1 & 0.00 & 0.00 & 0.00 & 4.30 & 0.00 & 0.00 \\
  & 
    & 2 & 0.12 & 0.00 & 0.70 & 3.90 & 12.60 & 0.00 \\
  & 
    & 4 & 0.13 & 0.00 & 1.60 & 28.90 & 45.20 & 1.70 \\
  & 
    & 8 & 0.88 & 7.25 & 1.20 & 25.00 & 77.00 & 40.40 \\
    \bottomrule
\end{tabular}
\end{table}

Table \ref{tab:vs_fi} demonstrates that variable‐selection errors are small under low noise ($\sigma=1$). Both FPR and FNR remain below $0.2\%$ for $n=125$ and effectively zero for $n=250$. As $\sigma$ increases, FNR rises sharply while FPR stays under $5\%$, indicating that the Step-One penalty ($\lambda_1$ and $\lambda_2$) grows more rapidly to prevent overfitting in high‐noise settings. 
For form identification, MoFI-optim yields a lower complex-to-simple misclassification rate ($r(1\!\to\!0)$) than MoFI-fix, whereas MoFI-fix achieves a lower simple-to-complex rate ($r(0\!\to\!1)$). This difference arises because cross‐validation selects smaller second‐stage penalties $\bar\lambda_1$ and $\bar\lambda_2$ in MoFI-optim than the fixed Step-One penalty used by MoFI-fix. Consequently, MoFI-fix is preferable when the goal is to maximize the detection of simple coefficient functions, while MoFI-optim is recommended to minimize false positives in simple‐form identification. 
In Scenarios I and II, both MoFI-fix and MoFI-optim deliver excellent form identification accuracy under low-to-moderate noise (with most misclassification rates remaining below 10\%) and remain robust even in high‐noise settings. In Scenario III, however, the weakened complementary components make form identification substantially more difficult.
Because MoFI-fix relies on a pre-fixed penalty, it cannot accommodate the intricate curvature, whereas MoFI-optim dynamically tunes to the true signal complexity, making it the superior choice.

\section{Application}
\label{sec:application}

We applied MoFI-FLR to the COG-BCI database \citep{hinss2023open}, a multi-session, multi-task electrophysiological (EEG) dataset designed for passive brain-computer interface research. The dataset includes recordings from 29 participants, each completing three sessions involving four distinct cognitive tasks intended to elicit varied mental states. In total, the database provides over 100 hours of publicly available EEG data.
For this study, we focused exclusively on the Psychomotor Vigilance Test (PVT), a widely used reaction-time task in cognitive neuroscience for assessing sustained attention and alertness \citep{basner2021response}. In the PVT, participants monitor a screen for a visual stimulus (typically a millisecond counter) presented at random intervals, and are instructed to press a button as quickly as possible upon stimulus onset. Performance is evaluated based on response times (RTs). Due to its high sensitivity to fatigue and minimal learning or aptitude effects, the PVT is considered a gold-standard tool for measuring neuro-behavioral consequences of sleep deprivation.
Each PVT session in the COG-BCI dataset contains 90 trials. Since fatigue accumulates progressively during the session, leading to increased RTs, we restricted our analysis to 30 early trials (i.e., trials $6$ to $35$) of each session to control for the late-stage fatigue effects.

During the PVT, EEG signals were recorded to investigate the neural correlates of attention and fatigue. To reduce noise and eliminate artifacts (such as those caused by blinks and eye movements), we applied the preprocessing pipeline proposed by \citet{yang2024learning} using EEGLAB \citep{delorme2004eeglab} in MATLAB. This pipeline includes bandpass filtering (1--40 Hz), removal of excessively noisy or malfunctioning channels, and independent component analysis (ICA) for artifact correction. 
These standard preprocessing steps help mitigate the impact of measurement noise in practice; see, for example, \citet{mumtaz2021review}. At the same time, to preserve potentially informative complex features in the EEG data, we adopted only a minimal preprocessing pipeline.

After preprocessing, 42 EEG channels were retained. The signals were sampled at $500$Hz, and task-related EEG segments spanning $0$ to $600$ms following stimulus onset were extracted. Trials containing high-amplitude artifacts or prolonged noisy segments were excluded. In addition, trials with abnormal RTs (defined as RTs greater than $700$ms or less than $100$ms) were removed.
The resulting dataset comprised $n = 1{,}765$ valid trials, each consisting of paired task EEG data and corresponding response time (RT) measurements.

The data exhibit a nested structure, with trials nested within sessions and sessions nested within subjects, and there may also be temporal dependence across trials.
To account for these sources of dependence, we analyze the session-normalized log response time, which removes subject-level and session-level mean effects. We further examine the resulting residuals and find no evidence of strong temporal dependence. This supports the use of our method for the normalized outcomes. Additional details are provided in Appendix~C.

\begin{figure}[htb!]
     \centering
     \begin{subfigure}[b]{0.57\textwidth}
         \centering
         \includegraphics[width=0.95\textwidth]{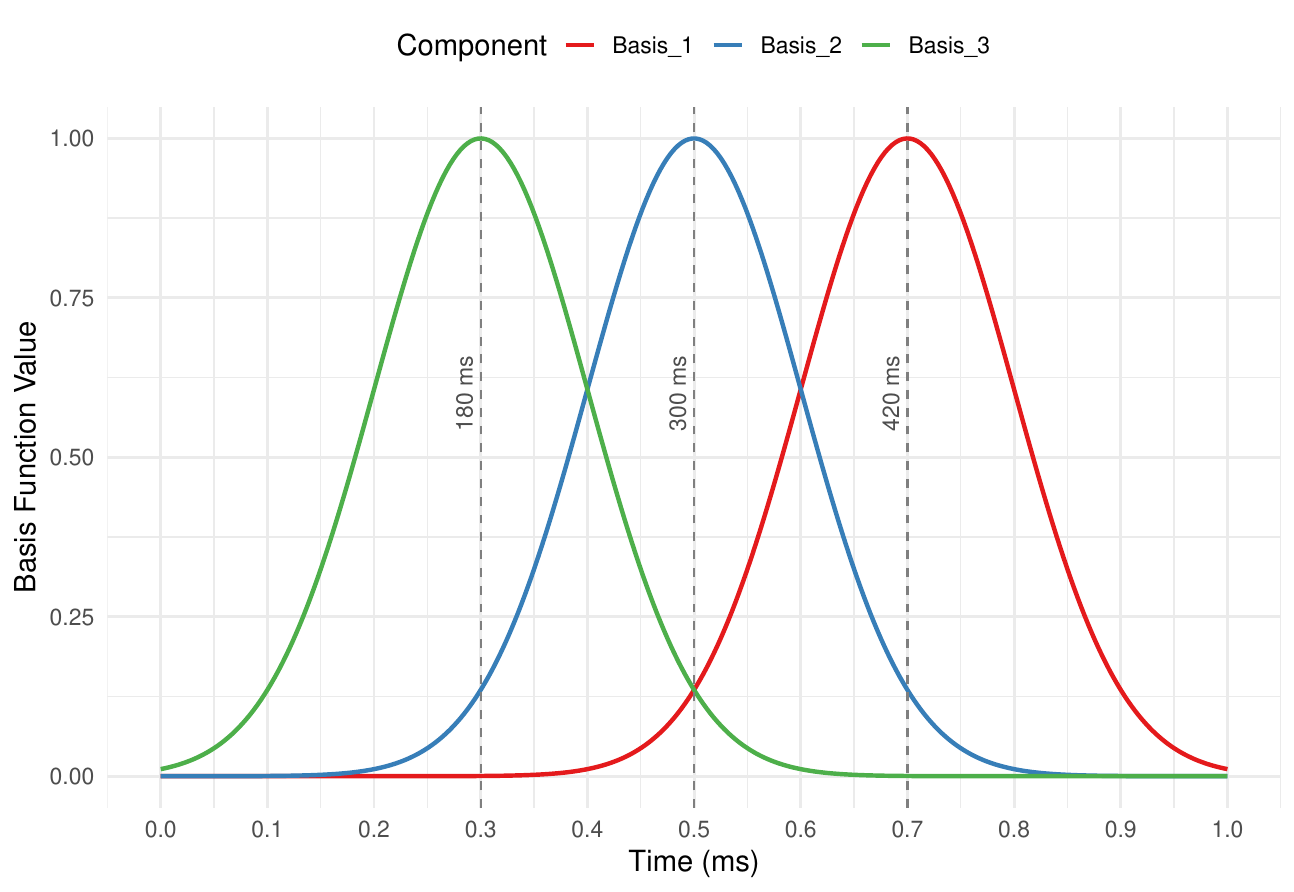}
         \caption{Three Gaussian basis functions}
     \end{subfigure}
     \hfill
     \begin{subfigure}[b]{0.42\textwidth}
         \centering
         \includegraphics[width=0.8\textwidth]{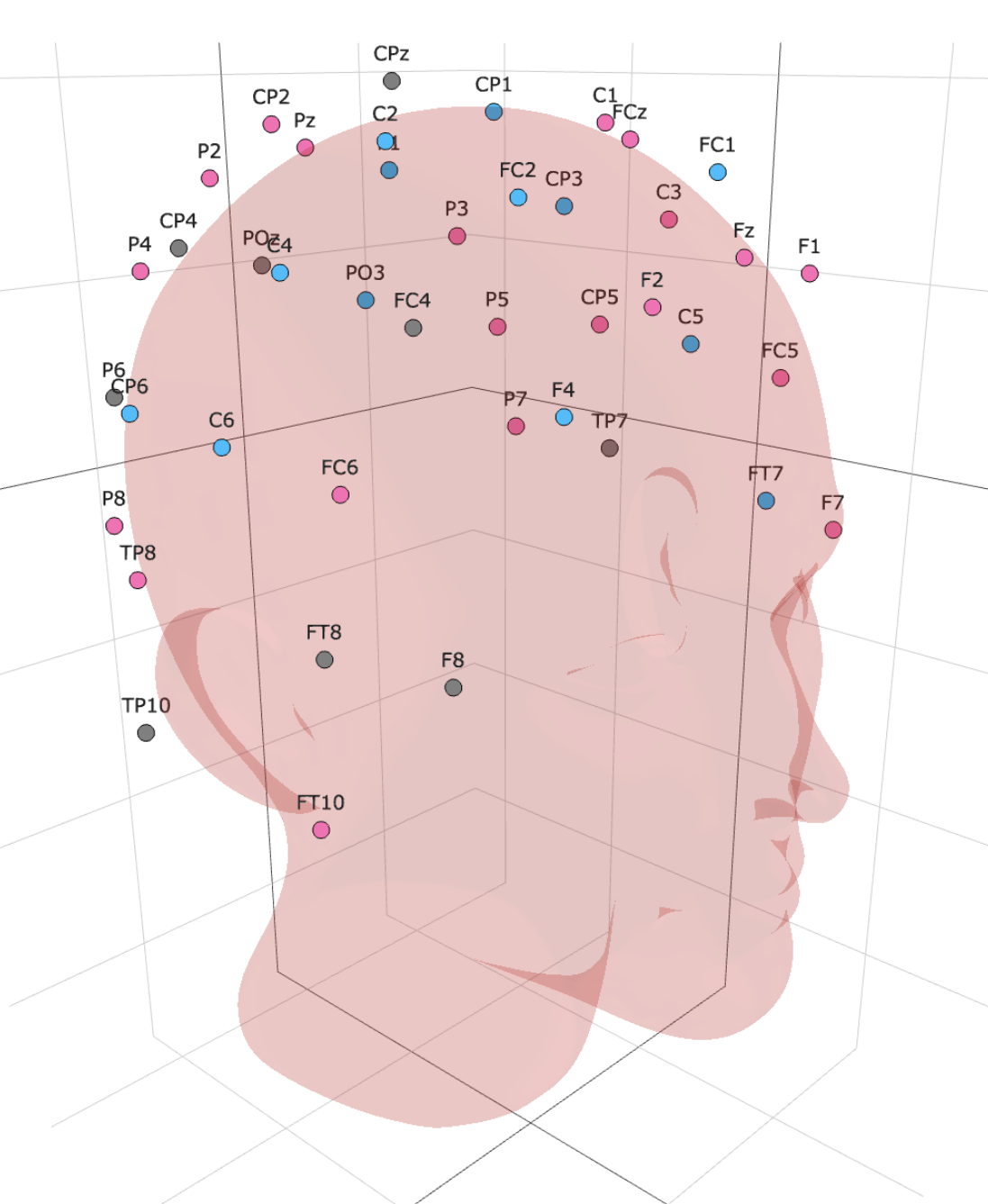}
         \caption{EEG channels on scalp surface}
     \end{subfigure}
     \begin{subfigure}[b]{0.65\textwidth}
         \centering
    \includegraphics[width=0.98\textwidth]{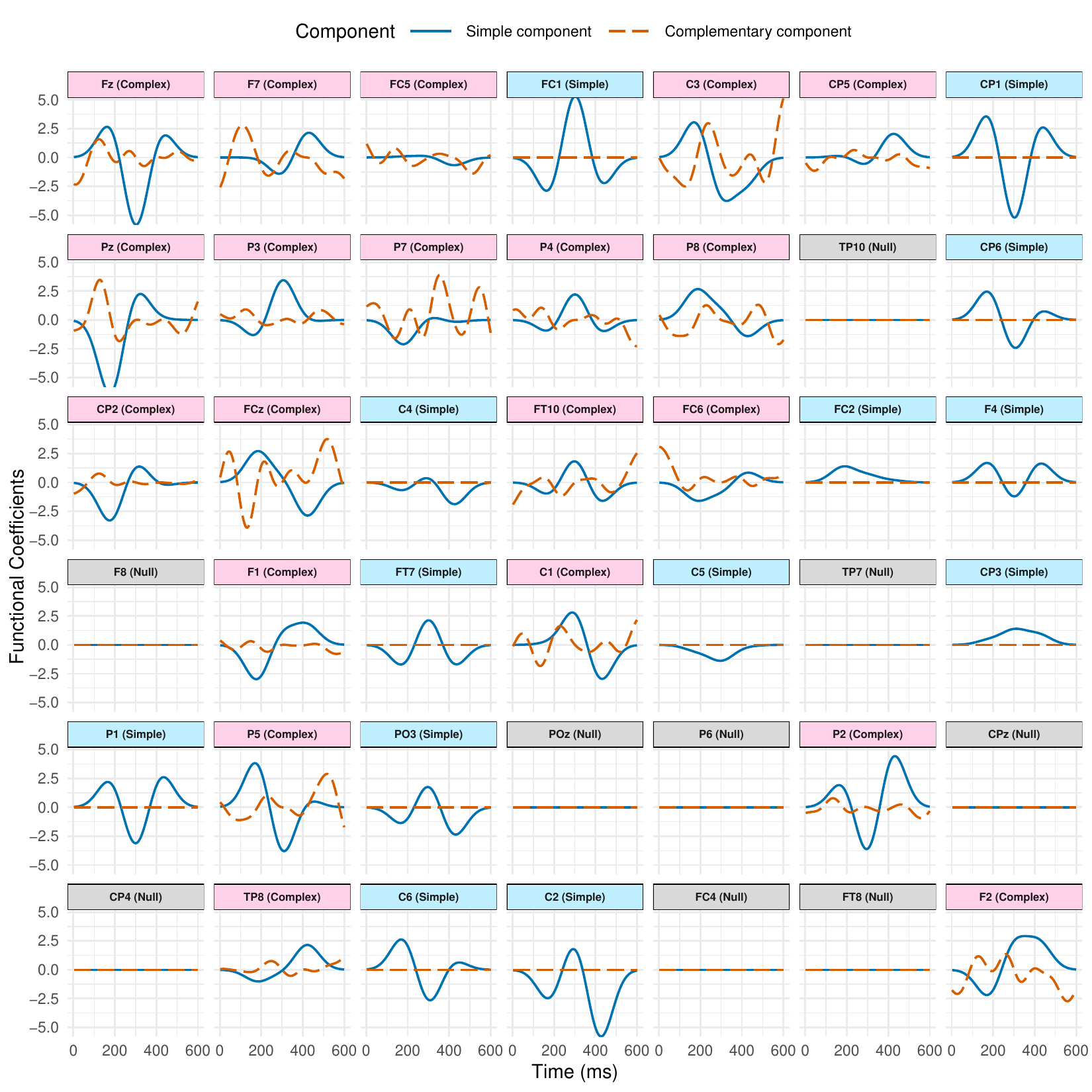}
         \caption{Estimated $\beta_j^{(0)}(t)$ and $\beta_j^{(1)}(t)$}
     \end{subfigure}
        \begin{subfigure}[b]{0.34\textwidth}
         \centering
    \includegraphics[width=0.88\textwidth]{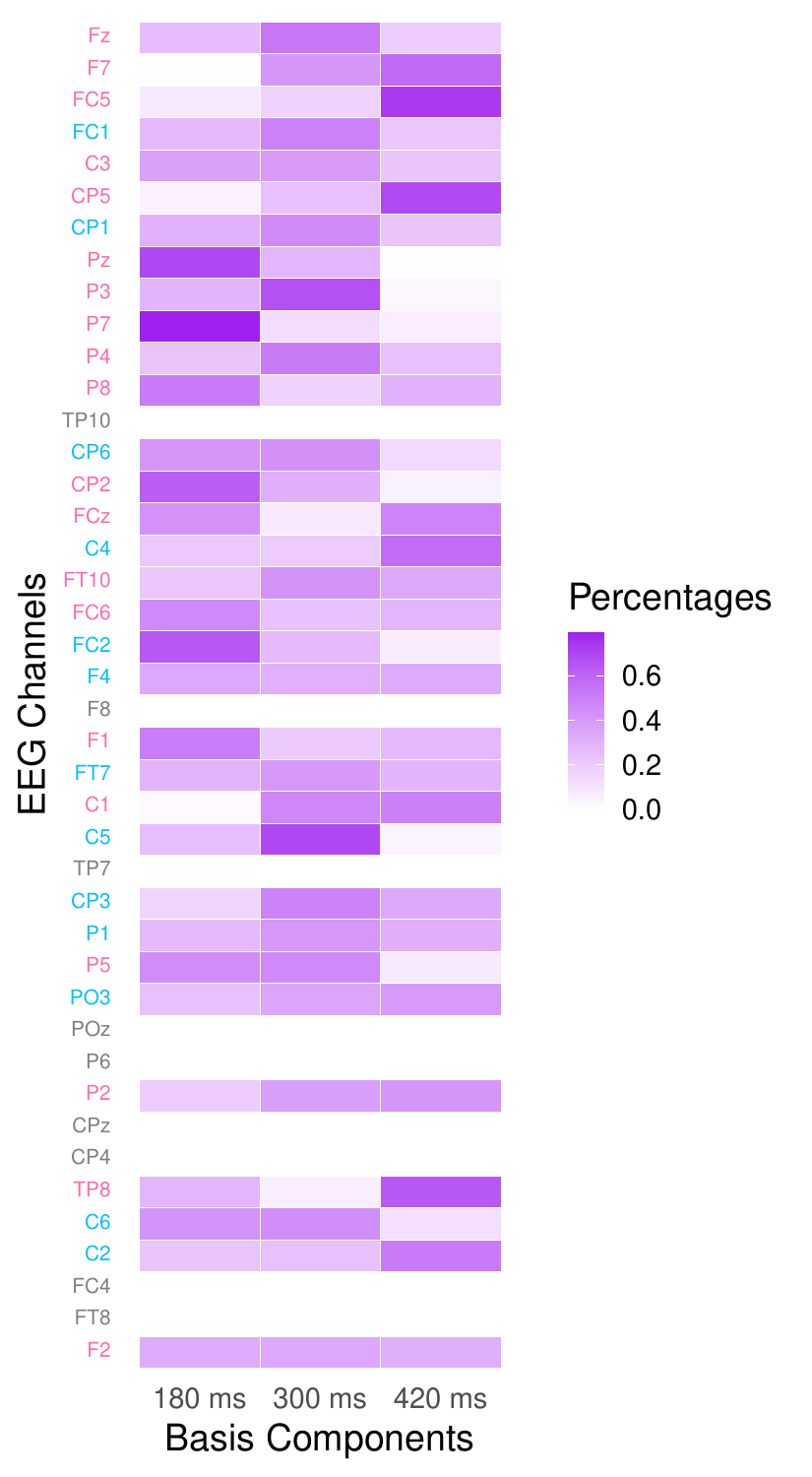}
         \caption{Simple coefficients percentages}
     \end{subfigure}
     \caption{Overview of the simple‐signal basis and functional‐coefficient fitting.}
     \label{fig:application1}
\end{figure}

We treat the multi-channel EEG data as high dimensional functional predictors and the session-normalized log response time as the outcome. To apply MoFI-FLR, we rescale the domain of the functional predictors to the interval $[0,1]$.
Next, we specify the same reproducing kernel Hilbert space (RKHS) for all functional coefficients, choosing the Gaussian radial basis function as the reproducing kernel:
$
K(s,t) = \exp\left\{ - 50 (s - t)^2 \right\}.
$
An event‐related potential (ERP) is a time‐locked EEG response to a specific sensory, cognitive, or motor event.
To capture the major ERP deflections, we define the ``simple" component of the functional coefficient as a linear combination of three Gaussian basis functions centered at $t_1 = 0.7$ ($420$ms; P3b), $t_2 = 0.5$ ($300$ms; P2/N2), and $t_3 = 0.3$ ($180$ms; P1/N1), reflecting canonical sensory and cognitive processing windows in visual oddball paradigms \citep{luck2014introduction}.
See Figure \ref{fig:application1}(a).
Let $\Phi_0(t) = \left( K(t_1, t), K(t_2, t), K(t_3, t) \right)^{\top}$  and define the $3 \times 3$ Gram matrix
$G = \left( K(t_{r_1}, t_{r_2}) \right)_{1 \le r_1, r_2 \le 3}$.
The kernel corresponding to the finite-dimensional subspace spanned by these basis functions is then given by $K_0(s, t) = \Phi_0^{\top}(s) G^{-1} \Phi_0(t)$, and the orthogonal complement kernel is defined as $K_1(s, t) = K(s, t) - K_0(s, t).$

We began by applying Step-One of our method to select relevant EEG channels. Using five-fold cross-validation to tune the regularization parameters, we found that 33 of the 42 channels were estimated to have nonzero coefficient functions.  The relatively large number of selected channels is likely due to the strong inter-channel dependence induced by volume conduction and spatial proximity (See Figure~\ref{fig:connectivity} in Appendix C). As a result, the selection outcome should be interpreted with caution.
In Step-Two, we again employed five-fold cross-validation to select the tuning parameters and identified 13 channels whose effects are well represented by the ``simple'' Gaussian basis structure. Figure~\ref{fig:application1}(b) shows the locations of these channels on the scalp. Figure~\ref{fig:application1}(c) presents their estimated coefficient functions, with the simple component \(\beta_j^{(0)}(t)\) and the complementary component \(\beta_j^{(1)}(t)\) displayed together in the same panel. Figure~\ref{fig:application1}(d) displays the normalized absolute loadings of these effects on the three Gaussian bases spanning the null RKHS.
Importantly, even channels identified as complex signals exhibit primary ERP deflections that align closely with our simple basis functions, demonstrating that this reduced representation captures the bulk of response‑time variability.
Because the simple component is spanned by three Gaussian basis functions, it already allows substantial flexibility and may therefore appear visually similar to the complementary component. 
However, a simple effect does not necessarily mean visually simple; rather, it means mathematically simple in the sense that it lies entirely in the prespecified finite-dimensional subspace. For example, in Figure~\ref{fig:application1}(c), the effects at channels FC1 and P1 are classified as simple because they are well captured by the three Gaussian basis functions.
By contrast, a complex effect typically reflects additional variation beyond these basis-driven patterns and therefore often appears more irregular; for example, channel C3 exhibits a substantial complementary component near the right boundary, around 600 ms, that cannot be explained by the three basis functions alone.

Our model exhibited strong predictive performance. In‐sample, it achieved a Pearson correlation of $0.76$ between observed and fitted RTs, with an RMSE $66\%$ that of a null predictor (relative RMSE $=0.66$). Out‐of‐sample (averaged over 100 random $80\%/20\%$ train–test splits), the mean Pearson correlation between true and predicted RTs was $0.64$, and the RMSE amounted to $78\%$ of the null model’s (relative RMSPE $=0.78$).
For comparison, the fully nonparametric baseline from Step-One achieved an out‐of‐sample correlation of $0.65$ and a relative RMSPE of $0.76$. Thus, our proposed approach matches the predictive accuracy of a fully nonparametric model while offering substantially greater interpretability across a large subset of channels.

\section{Discussion}
\label{sec:discussion}

In this paper, we present MoFI-FLR, a novel two-step estimation procedure for model-form identification in high-dimensional functional linear regressions. In the first step, we employ a functional elastic‐net penalty to screen out irrelevant functional covariates. 
In the second step, we apply an RKHS‐based orthogonal decomposition to the predictors retained in Step-One, splitting each coefficient into a finite‐dimensional (parametric) component and its infinite‐dimensional (nonparametric) complement. We then impose a penalty on the nonparametric component to perform model‐form identification.
Our non‐asymptotic analysis shows that, under mild regularity conditions and with suitably chosen tuning parameters, this procedure consistently recovers the correct set of active covariates and accurately identifies the form of functional coefficients.
The computational algorithm scales efficiently with sample size and the number of predictors, making it practical for modern high‐dimensional applications. Across an extensive suite of simulation studies, our method demonstrated superior performance in both variable selection and form selection, outperforming benchmark approaches that ignore functional structure and treat all effects as nonparametric. 
Our application of MoFI to the Psychomotor Vigilance Test (PVT) and its associated EEG recordings further demonstrates its ability to reveal meaningful brain–behavior relationships. In particular, we identified a small subset of EEG channels whose influence on reaction time is parsimoniously captured by three Gaussian basis functions with distinct peak latencies.

Compared with fully nonparametric alternatives, our two‐step method yields both more interpretable and more parsimonious estimates.
This clear decomposition allows practitioners to characterize each functional predictor with a minimal parametric component, augmented only by the essential nonparametric deviations.
The current theoretical framework assumes that the functional covariates are fully observed without measurement error. An important direction for future work is to extend the proposed method and its corresponding theory to settings where the functional predictors are contaminated by measurement error \citep{cardot2007smoothing}. Another practically relevant extension is to accommodate discretely observed functional covariates \citep{zhou2023functional}.

\acks{The R implementation of the MoFI-FLR algorithm is available in the GitHub repository at \url{https://github.com/xingcheg/MoFI-FLR/}}

\appendix

\section*{Appendix}

\subsection*{A. An illustrative Example for the Technical Conditions}

We now present an illustrative example of functional predictors that satisfy the technical conditions, specifically, Conditions \ref{ass:a2}, \ref{ass:a3}, \ref{ass:a4}, \ref{ass:a7}, \ref{ass:a8}.

We consider high-dimensional functional predictors with a partially separable covariance structure \citep{zapata2022partial}. This covariance structure captures the separability between the multivariate and functional aspects of the data, and has been widely used in functional data analysis. Particularly, we consider a simplified special case that
\begin{align}
   \mathbb{E}\bigl(X_{ij}\otimes X_{ij'}\bigr)=\sum_{m=1}^\infty a_m R_{m}^{(j,j')}\,\phi_m\otimes\phi_m, \label{eq:partial_separable}
\end{align}
where $a_m$ and $\phi_m$ are the $m$th eigenvalues and eigenfunction shared by all $X_j$'s, and $R_{m}^{(j,j')}$ is the correlation between the $m$th principal component score of $X_j$ and that of $X_{j'}$.
Assume further that $\mathbb{H}_j(K_j)=\mathbb{H}(K)$, for \(j=1,\dots,p\), and the covariance operator of \(X_j\) and the reproducing kernel \( K \) share the common orthonormal basis \(\{\phi_m\}_{m\ge 1}\), which is a commonly considered setting in functional linear regression \citep{yuan2010reproducing},
\begin{align*}
   K(s,t) = \sum_{m=1}^\infty \nu_m \phi_m(s) \phi_m(t). 
\end{align*}
Recall that $\wt X_{ij} = \mathscr K^{1/2} X_{ij}$, it then follows that 
\[
\mathscr{T}^{(j,j')}
=
\mathbb{E}\bigl(\widetilde X_{ij}\otimes \widetilde X_{ij'}\bigr)
=
\sum_{m=1}^\infty a_m \nu_m R_{m}^{(j,j')} \,\phi_m\otimes\phi_m.
\]

Let $\mathbf{R}_m=(R_{m}^{(j,j')})_{j,j'=1,\dots,p}$ and $\vartheta_m=a_m\nu_m$. 
If, in addition, \(\mathbf{R}_m\equiv \mathbf{R}\) for all \(m\), then the operator \(\mathscr{T}\) takes the form
\[
\mathscr{T}
=
\mathbf{R}\sum_{m=1}^\infty \vartheta_m\,\phi_m\otimes\phi_m.
\]
To incorporate dependence across predictors, a convenient choice is to let \(\mathbf{R}\) be an AR(1) correlation matrix, namely,
\[
R^{(j,j')}=\rho^{|j-j'|}, \qquad 1\le j,j'\le p,
\]
for some \(\rho\in[0,1)\).
We further assume that $\calS = \{1, \dots, q\}$, so that $\calS^c = \{q+1, \dots, p\}$.

\vskip5mm
\noindent \textbf{Check on Condition~\ref{ass:a2}:}
Since $\tr\bigl(\mathscr{T}^{(j,j)}\bigr)=
\sum_{m=1}^{\infty}\vartheta_m$, Condition~\ref{ass:a2} is easily satisfied if \(\sum_{m=1}^\infty \vartheta_m < \infty\). In particular, this holds whenever \(\vartheta_m\) decays at a rate faster than \( m^{-1} \).

\vskip5mm
\noindent \textbf{Check on Condition~\ref{ass:a3}:}
Using a result established in Section S.3.2 in \citet{guo2024rkhs}, we can show, under the covariance structure described above,
\begin{align}
    \varkappa(\lambda_2) = \vertiii{ \TSS (\TSS_{\lambda_2})^{-1} }_{\infty, \infty} \le 1+\frac{3 \rho}{1-\rho}, \label{equ:cond3_check1}
\end{align}
which does not depend on $\lambda_2$ and the true signal size $q$.
Next
\begin{align*}
    \vertiii{ \TScS (\TSS)^{-} }_{\infty, \infty} = \vertiii{ \mathbf{B} \sum_{m=1}^{\infty} \phi_m \otimes \phi_m }_{\infty, \infty} \le \max_{j \in\{1, \dots, p-q\}} \sum_{j'=1}^q |B^{(j,j')}|.
\end{align*}
where $\mathbf{B} = \mathbf{R}^{(\calS^c,\calS)} \left(\mathbf{R}^{(\calS,\calS)} \right)^{-1}$.
Under the AR(1) correlation structure 
we have
\[
\bigl(\mathbf R^{(\mathcal S,\mathcal S)}\bigr)^{-1}
=
\frac{1}{1-\rho^2}
\begin{pmatrix}
1 & -\rho & 0 & \cdots & 0\\
-\rho & 1+\rho^2 & -\rho & \cdots & 0\\
0 & -\rho & 1+\rho^2 & \ddots & \vdots\\
\vdots & \vdots & \ddots & 1+\rho^2 & -\rho\\
0 & 0 & \cdots & -\rho & 1
\end{pmatrix}.
\]
For any \(j\in\mathcal S^c\), namely \(j\ge q+1\),
\[
\mathbf R^{(j,\mathcal S)}
=
(\rho^{j-1},\rho^{j-2},\dots,\rho^{j-q}),
\]
and direct multiplication yields
\[
\left( B^{(j,1)}, \dots, B^{(j,q)} \right) = \mathbf R^{(j,\mathcal S)}
\bigl(\mathbf R^{(\mathcal S,\mathcal S)}\bigr)^{-1}
=
(0,\dots,0,\rho^{j-q}).
\]
Therefore
\begin{align}
    \vertiii{ \TScS (\TSS)^{-} }_{\infty, \infty} \le \rho. \label{equ:cond3_check2}
\end{align}
Combining Condition~3 with the requirement on \(\rho\) in \eqref{equ:cond3_check1} and \eqref{equ:cond3_check2}, we have
\begin{align*}
   \rho \left(1 + \frac{3\rho}{1-\rho} \right) \le \frac{C_{\min}}{C_{\max}}(1-\gamma).
\end{align*}
When \(\rho = 1/3\), the left-hand side is \(5/6\), so the above inequality can be satisfied when \(\gamma\) is sufficiently small. 

\vskip5mm
\noindent \textbf{Check on Condition~\ref{ass:a4}:}
Again, using a result established in Section S.3.2 of \citet{guo2024rkhs}, under the separable AR(1) covariance structure described above, 
\begin{align*}
    \aleph(\lambda_2)= \vertiii{ \left( \mathscr{T}^{(\calS, \calS)} - \mathscr{Q}^{(\calS, \calS)} \right)  \big(\mathscr{Q}^{(\calS, \calS)}_{\lambda_2}\big)^{-1} }_{\infty, \infty} < \frac{2\rho}{1-\rho},
\end{align*}
where the bound does not depend on \(\lambda_2\) or the true signal size \(q\). Consequently, a sufficient condition for the left-hand side to be smaller than \(1\) is $\rho < 1/3$.

\vskip5mm
\noindent \textbf{Check on Conditions~\ref{ass:a7} and~\ref{ass:a8}:}

Let us further assume that \(\mathbb{H}_{0j}(K_{0j})=\mathbb{H}_0(K_0)\), so that \(\mathbb{H}_{1j}(K_{1j})=\mathbb{H}_1(K_1)\). Therefore
\begin{align*}
   K_0(s,t) = \sum_{m=1}^{M_0} \nu_m \phi_m(s) \phi_m(t), \qquad K_1(s,t) = \sum_{m>M_0} \nu_m \phi_m(s) \phi_m(t).
\end{align*}
Recall that $\wt X_{ij}^{(1)} = \mathscr K_1^{1/2} X_{ij}$, it then follows that
\[
\mathscr{T}_1^{(j,j')}
:=
\mathbb{E}\bigl(\widetilde X_{ij}^{(1)}\otimes \widetilde X_{ij'}^{(1)}\bigr)
=R^{(j,j')} \sum_{m>M_0} \vartheta_m \,\phi_m\otimes\phi_m,
\]
and
\[
\widetilde X_{ij}^{(1)}(t)
=
\sum_{m>M_0}\nu_m^{1/2}Z_{ijm}\phi_m(t) \quad \mbox{with} \quad Z_{ijm} = \langle X_{ij}, \phi_m \rangle_2.
\]

Meanwhile, recall that $\boldsymbol{U}_{\cal W} = H_{\calS} \wt{X}_{\cal W}^{(1)}$, for ${\cal W} \in \{ {\cal S}_0, {\cal S}_1 \}$, where $H_{\calS} = I_n - \bZ_{\calS} (\bZ_{\calS}^{\top} \bZ_{\calS} )^{-1} \bZ_{\calS}^{\top}$, $\mathbf{Z}_{\calS} = (\mathbf{Z}_{ij}^{\top} )_{1\le i \le n}^{j \in \calS}$, $\mathbf{Z}_{ij} = (Z_{ij1}, \dots, Z_{ijM_0})^{\top}$.
We can show that, for \(m\le M_0\),
\begin{align*}
\mathbb{E}\!\left\{ Z_{ijm}\widetilde X_{ij'}^{(1)}(t)\right\}
&=
\sum_{m'>M_0}\nu_{m'}^{1/2}\phi_{m'}(t)\,
\mathbb{E}\!\left(Z_{ijm}Z_{ij'm'}\right) \\
&=
\sum_{m'>M_0} a_{m'}\nu_{m'}^{1/2} A_{m'}^{(j,j')} I(m=m')\,\phi_{m'}(t) \\
&= 0.
\end{align*}
Therefore, \(\widetilde X_{\mathcal W}^{(1)}\) and \(\bZ_{\mathcal S}\) are uncorrelated. Under the Gaussian assumption, uncorrelatedness implies independence, and hence \(\widetilde X_{\mathcal W}^{(1)}\) is independent of \(\bZ_{\mathcal S}\). It follows that \(\widetilde X_{\mathcal W}^{(1)}\) is also independent of \(H_{\mathcal S}\).

Define $A^{(\mathcal W_1,\mathcal W_2)}(s,t)$ as the kernel function for $\mathscr A^{(\mathcal W_1,\mathcal W_2)}$, $T_1^{(\mathcal W_1,\mathcal W_2)}(s,t)$ as the kernel function for $\mathscr{T}_1^{(\mathcal W_1,\mathcal W_2)}$. Then
\begin{align*}
A^{(\mathcal W_1,\mathcal W_2)}(s,t)
&=
\mathbb{E}\left\{
\frac{1}{n}\widetilde{X}_{\mathcal W_1}^{(1)\top}(s)
H_{\mathcal S}
\widetilde{X}_{\mathcal W_2}^{(1)}(t)
\right\} \\
&=
\mathbb{E}\left[
\mathbb{E}\left\{
\frac{1}{n}\widetilde{X}_{\mathcal W_1}^{(1)\top}(s)
H_{\mathcal S}
\widetilde{X}_{\mathcal W_2}^{(1)}(t)
\,\middle|\, H_{\mathcal S}
\right\}
\right] \\
&=
\mathbb{E}\left\{
\frac{1}{n}\operatorname{tr}(H_{\mathcal S})
\right\}
T_1^{(\mathcal W_1,\mathcal W_2)}(s,t) \\
&=
\frac{n-qM_0}{n}\,
T_1^{(\mathcal W_1,\mathcal W_2)}(s,t).
\end{align*}
As a result,
\begin{align}
    \mathscr A^{(\mathcal W_1,\mathcal W_2)}
=
\left(1-\frac{qM_0}{n}\right)
\mathscr T_1^{(\mathcal W_1,\mathcal W_2)}. \label{equ:A_special}
\end{align}

Hence, Conditions~\ref{ass:a7} and~\ref{ass:a8} may be equivalently formulated in terms of \(\mathscr T_1^{(\mathcal W_1,\mathcal W_2)}\) instead of \(\mathscr A^{(\mathcal W_1,\mathcal W_2)}\). These conditions are generally weaker than Conditions~\ref{ass:a3} and~\ref{ass:a4} imposed on \(\mathscr T^{(\mathcal W_1,\mathcal W_2)}\), because \(\mathscr T^{(\mathcal W_1,\mathcal W_2)}\) includes both the simple component (\(m=1,\dots,M_0\)) and the complementary component (\(m>M_0\)), whereas \(\mathscr T_1^{(\mathcal W_1,\mathcal W_2)}\) involves only the complementary component.

\subsection*{B. Technical Proofs}

\paragraph{B.1. Proof of the Proposition \ref{lemma:solution_form}}

    Consider the objective function \eqref{equ:mini1} in Step 1:
    \begin{align*}
    \ell(\boldsymbol{f}):=\frac{1}{2n}\sum_{i=1}^n
    \left(Y_i-\sum_{j=1}^p \langle \widetilde{X}_{ij}, f_j\rangle_2\right)^2+\lambda_1 \sum_{j=1}^p \|\Psi_j f_j\|_2+\frac{\lambda_2}{2}\sum_{j=1}^p \|f_j\|_2^2.
    \end{align*}
    For any minimizer $\widetilde f_j$ admits the decomposition
    \[\widetilde f_j=\widehat f_j+\eta_j,\]
    where
    \[
    \widehat f_j(\cdot)=\sum_{i=1}^n c_{ij}\widetilde X_{ij}(\cdot)\in \mathbb M_{nj},
    \qquad
    \eta_j(\cdot)\in \mathbb M_{nj}^{\perp}.
    \]
    It then follows that
    \begin{align*}
        \langle \widetilde X_{ij},\widetilde f_j\rangle_2
    =\langle \widetilde X_{ij},\widehat f_j\rangle_2, \qquad \|\widetilde f_j\|_2^2=\|\widehat f_j\|_2^2+\|\eta_j\|_2^2,
    \end{align*}
    and
    \begin{align*}
        \|\Psi_j\widetilde f_j\|_2^2=\|\Psi_j\widehat f_j\|_2^2+\|\Psi_j\eta_j\|_2^2.
    \end{align*}

The last identity follows from Condition~\ref{ass:a1}. Indeed, by self-adjointness of \(\Psi_j\),
\[
\langle \Psi_j \widehat f_j,\, \Psi_j \eta_j\rangle_2
=
\langle \Psi_j^2 \widehat f_j,\, \eta_j\rangle_2.
\]
Since $\Psi_j^2 \widehat f_j \in \mathbb M_{nj}$ and $\eta_j \in \mathbb M_{nj}^\perp$, the inner product is zero.
Therefore, replacing $\widetilde f_j$ by $\widehat f_j$ leaves the data-fitting term unchanged while not increasing the penalty terms. Hence, any minimizer must satisfy $\eta_j\equiv 0$, so that $\widetilde f_j\in \mathbb M_{nj}$.
A similar argument applies to the minimization problem in \eqref{equ:mini2} for Step~2 under Condition~\ref{ass:a6}. Therefore, we also have $\widetilde f_j^{(1)}\in \mathbb M_{nj}^{(1)}$.

\paragraph{B.2. Proof of the Theorem \ref{thm:2}}

The proof of Theorem \ref{thm:2} can be divided into two parts.  In the first part, we show that 
$\widehat{\mathcal S} = \mathcal S$
with probability tending to one.  Indeed, by Theorem 1 of \cite{guo2024rkhs}, under Conditions \ref{ass:a1}–\ref{ass:a5} and the bounds on $\lambda_1$ and $\lambda_2$ in \eqref{equ:bar_lambda_12_bound_1}, one has
\begin{align*}
    \Pr\bigl(\widehat{\mathcal S} = \mathcal S\bigr)
  \ge 1 - P_1,
  \quad
  P_1 = \exp \bigl(-D\lambda_{2}^{2} n / q \bigr),
\end{align*}
where $D$ is the constant defined in \eqref{equ:D_form} in Theorem \ref{thm:2}.

In the second part, we only need to show conditional form‐selection consistency, that is:
\begin{align*}
    \Pr\bigl(\widehat\calS_1 = \calS_1 \mid \widehat\calS = \calS\bigr) > 1 - P_2, \quad
  P_2 = \exp \bigl(-\bar{D} \bar\lambda_{2}^{2} n / q_1 \bigr),
\end{align*}
with $\bar{D}$ defined in \eqref{equ:D_form}.

We further split the remainder of the proof into two parts.  
In Part 1, we bound $\Pr\bigl(\widehat\calS_1 \subset \calS_1 \mid \widehat\calS = \calS\bigr)$
and in Part 2, we bound $\Pr\bigl(\widehat\calS_1 \supset \calS_1 \mid \widehat\calS = \calS\bigr)$.
Then
\begin{align*}
    \Pr\bigl(\widehat\calS_1 \neq \calS_1 \mid \widehat\calS = \calS\bigr) < \Pr\bigl(\widehat\calS_1 \not\subset \calS_1 \mid \widehat\calS = \calS\bigr) + \Pr\bigl(\widehat\calS_1 \not\supset \calS_1 \mid \widehat\calS = \calS\bigr).
\end{align*}

\vspace{1em}

\noindent \textbf{Part 1: bound $\Pr\bigl(\widehat\calS_1 \subset \calS_1 \mid \widehat\calS = \calS\bigr)$}

Recall that $\mathbf{Z}_{\calS} = (\mathbf{Z}_{ij}^{\top} )_{1\le i \le n}^{j \in \calS}$,  $\wt{X}_{\calS}^{(1)} = ( \wt{X}_{ij}^{(1)})_{1\le i \le n}^{j \in \calS}$. Denote $\ba_{\calS} = \left( \ba_j^{\top} \right)_{j \in \calS}^{\top}$ to be a vector of length-$qM_0$,
From \eqref{equ:mini2}, 
we can easily derive the ordinary least square (OLS) estimator of $\ba_{\calS}$ by first assuming that $\boldsymbol{f}_{\calS}^{(1)}$ is given, that is
\begin{align*}
    \wh{\ba}_{\calS} = \left(\bZ_{\calS}^{\top} \bZ_{\calS} \right)^{-1} \bZ_{\calS}^{\top} \left( \bY_n - \int_{0}^1 \wt{X}_{\calS}^{(1)}(t) \boldsymbol{f}_{\calS}^{(1)}(t) dt  \right).
\end{align*}
Plug-in the OLS estimator back to \eqref{equ:mini2}, we have
\begin{align} \label{equ:mini2_1}
\ell(\boldsymbol{f}_{\calS}^{(1)})
&= \frac{1}{2} \left\langle  \mathscr{A}_n^{(\calS, \calS)} (\boldsymbol{f}_{\calS}^{(1)} - {\boldsymbol{f}_{0\calS}^{(1)}}),  \ \boldsymbol{f}_{\calS}^{(1)} - {\boldsymbol{f}_{0\calS}^{(1)}} \right\rangle_{2} - \left\langle \boldsymbol{r}_n,  \ \boldsymbol{f}_{\calS}^{(1)} - {\boldsymbol{f}_{0\calS}^{(1)}} \right\rangle_{2} \nonumber\\
& \quad + \frac{\bar\lambda_2}{2} \left\| \boldsymbol{f}_{\calS}^{(1)} \right\|_{2}^2 + \bar\lambda_1\sum_{j=1}^p \left\| \Psi_j^{(1)} {f_j^{(1)}} \right\|_{2} 
+ \frac{1}{2n} \left\| H_{\calS}  {\varepsilon}_n \right\|_2^2 ,
\end{align}
where $H_{\calS} = I_n - \bZ_{\calS}\left(\bZ_{\calS}^{\top} \bZ_{\calS} \right)^{-1} \bZ_{\calS}^{\top}$. 
Recall that $\boldsymbol{U}_{\cal S} = H_{\calS} \wt{X}_{\cal S}^{(1)}$, then 
$\boldsymbol{r}_n = n^{-1} \boldsymbol{U}_{\cal S}^{\top} \boldsymbol{\varepsilon}_n$.

Next, we present the KKT condition for the optimization problem \eqref{equ:mini2_1}, which will play a central role in the proof of Theorem \ref{thm:2}.
\begin{proposition}
\label{theorem:kkt}
Suppose Condition \ref{ass:a6} holds. Then, for all $\bar\lambda_1, \bar\lambda_2 > 0$, the solution $\wh{\boldsymbol{f}}_{\calS}^{(1)}$ for \eqref{equ:mini2}
exists uniquely and satisfies 
\begin{align}
\label{equ:kkt}
	 \mathscr{A}_n^{(\calS, \calS)}( \wh{\boldsymbol{f}}_{\calS}^{(1)} - {\boldsymbol{f}_{0\calS}^{(1)}} ) -  \boldsymbol{r}_n +     \bar\lambda_2 \wh{\boldsymbol{f}}_{\calS}^{(1)} +    \bar\lambda_1  \boldsymbol{\omega}_{\calS} = 0,
\end{align}
where $\boldsymbol{\omega}_{\calS} = (\omega_j)_{j \in \calS}^{\top}$ with 
\begin{align*}
    \omega_j = \frac{(\Psi_j^{(1)})^2 \wh{f}_j^{(1)} } { \left\| \Psi_j^{(1)} \wh{f}_j^{(1)} \right\|_2}
\end{align*}
if $\wh{f}_j^{(1)} \not =0$ and $\omega_j = \Psi_j^{(1)} \eta_j$ for some  $\eta_j$ with $\|\eta_j\|_2\le 1$ if $\wh{f}_j^{(1)} = 0$.
\end{proposition}

Write ${\calS} = ({\calS_1}, {\calS_0})$ by grouping the signal set $\calS$ in complex signal set $\calS_1$ and simple signal set $\calS_0$. 
We assume that $\boldsymbol{f}_{0 \calS_0}^{(1)} = \boldsymbol{0}$, and we have $\boldsymbol{f}_{0 \calS}^{(1)}=(\boldsymbol{f}_{0 \calS_1}^{(1) \top}, {\pmb 0}^\top)^\top$. Similarly, partition $\wh{\boldsymbol{f}}_{\calS}^{(1)}= (\wh{\boldsymbol{f}}_{\calS_1}^{(1) \top}, \wh{\boldsymbol{f}}_{\calS_0}^{(1) \top})^\top$, $\boldsymbol{r}_n=(\boldsymbol{r}_{\calS_1}^\top, \boldsymbol{r}_{\calS_0}^\top)^\top$ and $\bdomega=(\bdomega_{\calS_1}^\top, \bdomega_{\calS_0}^\top)^\top$.
With the partitions defined above, the KKT condition in (\ref{equ:kkt}) can be rewritten as
\begin{align}
\label{equ:kkt0}
\left(
\begin{matrix} 
\Anll & \Anlo \\
\Anol & \Anoo
\end{matrix}
\right)
 \left(
\begin{matrix} 
\wh{\boldsymbol{f}}_{\calS_1}^{(1)} - \boldsymbol{f}_{0\calS_1}^{(1)}  \\
\wh{\boldsymbol{f}}_{\calS_0}^{(1)}
\end{matrix}
\right) - 
 \left(
\begin{matrix} 
\boldsymbol{r}_{\calS_1}  \\
\boldsymbol{r}_{\calS_0}
\end{matrix}
\right) 
+ \bar\lambda_2  \left(
\begin{matrix} 
\wh{\boldsymbol{f}}_{\calS_1}^{(1)}  \\
\wh{\boldsymbol{f}}_{\calS_0}^{(1)}
\end{matrix}
\right) + \bar\lambda_1
 \left(
\begin{matrix} 
\bdomega_{\calS_1}  \\
\bdomega_{\calS_0}
\end{matrix}
\right) = \boldsymbol{0}.
\end{align}
To utilize the Primal-Dual Witness argument in \cite{wainwright2009sharp}, let $\check{\boldsymbol{f}}_{\calS_1}^{(1)}$ be the solution of \eqref{equ:mini2} knowing the true complex set $\calS_1$. In other words, $\check{\boldsymbol{f}}_{\calS_1}^{(1)}$ is the value of $\boldsymbol{f}_{\calS_1}^{(1)}$ that minimizes
\begin{align*}
    \frac{1}{2} 
 \left\langle  \Anll(\boldsymbol{f}_{\calS_1}^{(1)}-\boldsymbol{f}_{0\calS_1}^{(1)}),  \boldsymbol{f}_{\calS_1}^{(1)}-\boldsymbol{f}_{0\calS_1}^{(1)}  \right\rangle_{2} -  \left\langle  \boldsymbol{r}_{\calS_1} , \boldsymbol{f}_{\calS_1}^{(1)}-\boldsymbol{f}_{0\calS_1}^{(1)} \right\rangle_{2} + \sum_{j\in \calS_1} \mathrm{Pen}\left(f_j^{(1)}; \bar\lambda_1, \bar\lambda_2 \right).
\end{align*}
Using similar arguments as for Proposition \ref{theorem:kkt}, we have
\begin{align}
	\Anll( \check{\boldsymbol{f}}_{\calS_1}^{(1)}-\boldsymbol{f}_{0\calS_1}^{(1)}) -  \boldsymbol{r}_{\calS_1} + \bar\lambda_2 \check{\boldsymbol{f}}_{\calS_1}^{(1)} + \bar\lambda_1  \bdomega_{\calS_1} = \boldsymbol{0},
 \label{equ:kkt1}
\end{align}
where $\bdomega_{\calS_1} = (\Psi_j^{(1)} \eta_j,j\in \calS_1)$.
By Proposition \ref{theorem:kkt}, if we can show that $ \left(\check{\boldsymbol{f}}_{\calS_1}^{(1) \top}, \boldsymbol{0}^\top\right)^{\top}$ solves \eqref{equ:kkt0}, then $\wh{\boldsymbol{f}}_{\calS}^{(1)} = \left(\check{\boldsymbol{f}}_{\calS_1}^{(1) \top}, \boldsymbol{0}^\top\right)^{\top}$ and $\wh{\calS}_1 \subset \calS_1$. It remains to show
\begin{align}
	& \Anol( \check{\boldsymbol{f}}_{\calS_1}^{(1)}-\boldsymbol{f}_{0\calS_1}^{(1)}) -  \boldsymbol{r}_{\calS_0} + \bar\lambda_1  \bdomega_{\calS_0} = \boldsymbol{0},  \label{equ:kkt2}
\end{align}
for some $\bdomega_{\calS_0}$ satisfying $\bdomega_{\calS_0}=(\Psi_j^{(1)} \eta_j, j\in \calS_0)$ where $\|\boldsymbol{\eta}_{\mathcal{S}_0} \|_{\infty}\le 1$.  
However, by (\ref{equ:kkt1}), 
\begin{align}
	\check{\boldsymbol{f}}_{\calS_1}^{(1)}-\boldsymbol{f}_{0\calS_1}^{(1)} =   \left(\dAnll\right)^{-1}  \left( \boldsymbol{r}_{\calS_1} - \bar\lambda_2 \boldsymbol{f}_{0\calS_1}^{(1)} - \bar\lambda_1  \bdomega_{\calS_1} \right),
 \label{equ:f_diff}
\end{align}
and hence, upon combining (\ref{equ:kkt2}) and (\ref{equ:f_diff}), any $\wsc$ that solves \eqref{equ:kkt2} must satisfy
\begin{align}\label{equ:omega_S_c1}
\begin{split}
	\bdomega_{\calS_0}  &:= \bar{\lambda}_1^{-1} \left\{ \boldsymbol{r}_{\calS_0} - \Anol  \left(\dAnll\right)^{-1} \boldsymbol{r}_{\calS_1} \right\} \\
& \hspace{.6cm} + \Anol  \left(\dAnll\right)^{-1} \left(\bar\lambda_2 \bar{\lambda}_1^{-1} \boldsymbol{f}_{0\calS_1}^{(1)} + \bdomega_{\calS_1} \right). 
\end{split}
\end{align}
By Condition \ref{ass:a6}, the existence of $\bdomega_{\calS_0}$ satisfying \eqref{equ:kkt2} is guaranteed by
\begin{align} \label{e:thm1_i}
	\| \bdomega_{\calS_0} \|_{\infty} \le C_{\min}.
\end{align}
The rest of the proof will be focusing on \eqref{e:thm1_i}.

Note that the first term on the right-hand side of \eqref{equ:omega_S_c1} can be rewritten as
\begin{align*}
    \frac{\sigma}{n \bar\lambda_1 } \boldsymbol{U}_{\calS_0}^{\top} \left( \mathbf{I}_n - \boldsymbol{\Delta}_n  \right)  \boldsymbol{z}_n, \quad \mbox{where} \quad
    \boldsymbol{\Delta}_n =  n^{-1} \int \boldsymbol{U}_{\calS_1}(t) \left\{\left(\dAnll\right)^{-1} \boldsymbol{U}_{\calS_1}^{\top} \right\} (t) dt,
\end{align*}
$\boldsymbol{z}_n = \boldsymbol{\varepsilon}_n / \sigma$.
If $\widehat{\calS_1} \not\subset \calS_1$ then \eqref{e:thm1_i} fails.

Let $\bar\gamma$ be as in Condition \ref{ass:a7}. Suppose $\bar\lambda_1 >  D_1^* (\sigma+1) \tau^{1/2}  (C_{\min} \bar\gamma)^{-1} \left\{\log(q_0) / n \right\}^{1/2}$ for some constant $D_1^*$.
We have
\begin{align}
	& \pr\left(\bigg\| \frac{\sigma}{n \bar\lambda_1 } \boldsymbol{U}_{\calS_0}^{\top} \left( \mathbf{I}_n - \boldsymbol{\Delta}_n  \right)  \boldsymbol{z}_n  \bigg\|_{\infty} \ge \frac{ \bar\gamma C_{\min}}{9}\right) \le \exp\left( - D^{(1)} \bar\lambda_1^2 n  \right) \label{equ:lm:(i)}
\end{align}
where $D^{(1)} = D_2^* C_{\min}^2 \bar\gamma^2 (\sigma + 1)^{-2} \tau^{-1}$ and $D_2^*$ is a universal constant.
The proof of \eqref{equ:lm:(i)} follows by analogy with Lemma 2 of \cite{guo2024rkhs}, and is therefore omitted.

On the other hand, for some constant $D_3^*$,
\begin{align*}
    \bar\lambda_2 > D_3^* \frac{\tau (\bar\rho_1 + 1 )}{(C_{\min}/C_{\max})^2 \bar\gamma^2}  \max\left({q_1\log(q_0)\over n},\sqrt{q_1^2\over n} \right) 
    \quad \mbox{and} \quad \bar\lambda_1 / \bar\lambda_2 > \left({3\over\bar\gamma} - 2 \right) C_{\max}^{-1}.
\end{align*}
Then
\begin{align} 
	& \pr\left\{   \left\| \Anol  \left(\dAnll\right)^{-1} \left(\bar\lambda_2 \bar{\lambda}_1^{-1} \boldsymbol{f}_{0\calS_1}^{(1)} + \bdomega_{\calS_1} \right) \right\|_{\infty}
 \ge \left(1-{2 \bar\gamma \over 9}\right) C_{\min} \right\} \nonumber \\
 &\quad \le \exp\left( - D^{(2)} {\bar\lambda_2^2 n \over q_1} \right) \label{equ:lm:(ii)}
\end{align}
where $D^{(2)} = D_4^*  (C_{\min}/C_{\max})^2 \bar\gamma^2 (\bar\rho_1+1)^{-2} \tau^{-1}$ and $D_4^*$ is a universal constant.
The proof of \eqref{equ:lm:(ii)} closely parallels that of Lemma 3 in \cite{guo2024rkhs}.

As a result of \eqref{equ:lm:(i)} and \eqref{equ:lm:(ii)}, 
\begin{align} 
\begin{split}
\mathbb{P} \left( \widehat{\calS}_1 \not\subset \calS_1 \mid  \widehat{\calS} = \calS \right) 
& \le \mathbb{P} \left( \|  \bdomega_{\calS_0} \|_{\infty} > \left(1 - \frac{\bar\gamma}{9}\right)C_{\min} \right) \\
& \le \exp\left( - D^{(1)} \bar\lambda_1^2 n  \right)
+\exp\left( - D^{(2)} { \bar\lambda_2^2 n \over q_1} \right).
\end{split}
    \label{equ:tail_bound_final_1}
\end{align}

Note that $\exp\left( - D^{(1)} \bar\lambda_1^2 n  \right) \le \exp\left( - D^{(1)} C_{\max}^{-2} q_1 ( \bar\lambda_2^2 n / q_1)  \right)$ since $\bar\lambda_1> C_{\max}^{-1} \bar\lambda_2$. It can be shown that the rhs of \eqref{equ:tail_bound_final_1} can be bounded by the probability in \eqref{equ:vs_consistency_rate1} under Condition \eqref{equ:bar_lambda_12_bound_1}.

\vspace{1em}
\noindent \textbf{Part 2: bound $\Pr\bigl(\widehat\calS_1 \supset \calS_1 \mid \widehat\calS = \calS\bigr)$}

Next, we need to show that $\| \widehat{f}_j^{(1)} \|_{2} > 0$ for all $j \in \calS_1$ with the probability lower bound stated in the theorem. 
Note that 
\begin{align*}
    &\pr(\wh \calS_1 \supset \calS_1 \mid \widehat\calS = \calS) \\
    &= \pr \left( \min_{ j \in \calS_1} \| \widehat{f}_j^{(1)} \|_{2} > 0 \mid \widehat\calS = \calS \right) \\
    &\ge \pr \left(  \min_{j\in \calS_1} \| (\mathscr{A}^{(j,j)} )^{1/2} \widehat{f}_j^{(1)} \|_{2} > 0 \mid \widehat\calS = \calS \right).
\end{align*}
By the triangle inequality,
\begin{align*}
	\min_{j \in \calS_1} \left\| (\mathscr{A}^{(j,j)} )^{1/2} \wh{f}_j^{(1)} \right\|_2 
	& \ge \min_{j \in \calS_1} \left\| (\mathscr{A}^{(j,j)} )^{1/2} f_{0j}^{(1)} \right\|_2 -  \left\| (\mathscr{B}^{(\calS_1, \calS_1)} )^{1/2} (\wh{\boldsymbol{f}}_{\calS_1}^{(1)}-\boldsymbol{f}_{0\calS_1}^{(1)}) \right\|_{\infty}.
\end{align*}
Also, by \eqref{equ:f_diff}, we have
\begin{align*}
	\wh{\boldsymbol{f}}_{\calS_1}^{(1)}-\boldsymbol{f}_{0\calS_1}^{(1)} =&  \left(\All_{\bar\lambda_2}\right)^{-1}  \left( \boldsymbol{r}_{\calS_1} - \bar\lambda_2 \boldsymbol{f}_{0\calS_1}^{(1)} - \bar\lambda_1  \bdomega_{\calS_1} \right) \\
	& + \bigg\{ (\All_{n,\bar\lambda_2})^{-1} - (\All_{\bar\lambda_2})^{-1} \bigg\} \left( \boldsymbol{r}_{\calS_1} - \bar\lambda_2 \boldsymbol{f}_{0\calS_1}^{(1)} - \bar\lambda_1  \bdomega_{\calS_1} \right).
\end{align*}
Since
$(\All_{_{n,\bar\lambda_2}})^{-1} - (\All_{\bar\lambda_2})^{-1} = (\All_{\bar\lambda_2})^{-1} \left(\All - \Anll \right)(\All_{n,\bar\lambda_2})^{-1}$,
\begin{align*}
	& \left\| (\mathscr{B}^{(\calS_1, \calS_1)} )^{1/2} (\wh{\boldsymbol{f}}_{\calS_1}^{(1)}-\boldsymbol{f}_{0\calS_1}^{(1)}) \right\|_\infty \\
 & \le  \vertiii{ (\mathscr{B}^{(\calS_1, \calS_1)} )^{1/2}  (\All_{\bar\lambda_2})^{-1}}_{\infty,\infty} \left\{ \left\| \boldsymbol{r}_{\calS_1} \right\|_{\infty} + \bar\lambda_2 \left( \| \boldsymbol{f}_{0\calS_1}^{(1)} \|_{\infty} + \frac{\bar\lambda_1}{\bar\lambda_2} C_{\max} \right) \right\} \\
    & \hspace{.5cm} \times \left( 1 + \frac{\sqrt{q_1}}{\bar\lambda_2} \vertiii{\All - \Anll}_{2,2} \right),
\end{align*}
where we applied the inequality 
\begin{align*}
\vertiii{ \left(\All - \Anll \right)(\All_{n,\bar\lambda_2})^{-1} }_{\infty,\infty}  \le \frac{\sqrt{q_1}}{\bar\lambda_2} \vertiii{ \All - \Anll }_{2,2}.
\end{align*}
Also, under Condition \ref{ass:a8}, for any $\bar\lambda_2 > 0$
\begin{align}
\vertiii{ (\mathscr{B}^{(\calS_1, \calS_1)} )^{1/2}  (\All_{\bar\lambda_2})^{-1}}_{\infty,\infty} < \frac{6-4\bar\aleph(\bar\lambda_2)}{1-\bar\aleph(\bar\lambda_2)} \frac{1}{\sqrt{\bar\lambda_2}}.
\label{equ:bound_4}
\end{align}
The proof of \eqref{equ:bound_4} can be found in Lemma 4 of \cite{guo2024rkhs}.

By \eqref{equ:bound_4} and the fact that $ \| \boldsymbol{f}_{0\calS_1}^{(1)} \|_{\infty} \le \| \fzeros \|_{\infty} = 1$, 
\begin{eqnarray}\label{equ:f_norm_diff}
    && \left\| (\mathscr{B}^{(\calS_1, \calS_1)} )^{1/2} (\wh{\boldsymbol{f}}_{\calS_1}^{(1)}-\boldsymbol{f}_{0\calS_1}^{(1)}) \right\|_\infty \\
&& \le \frac{6-4\bar\aleph(\bar\lambda_2)}{(1-\bar\aleph(\bar\lambda_2))\sqrt{\bar\lambda_2}} \left\{ \left\| \boldsymbol{r}_{\calS_1} \right\|_{\infty} + \bar\lambda_2   + \bar\lambda_1 C_{\max}  \right\} 
    \left( 1 + \frac{\sqrt{q_1}}{\bar\lambda_2} \vertiii{\All - \Anll}_{2,2} \right), \nonumber
\end{eqnarray}
Thus, with $\bar{G}$ as given in Condition \ref{ass:a9},
\begin{align}
\begin{split}
    & \pr\left( \left\| (\mathscr{B}^{(\calS_1, \calS_1)} )^{1/2} (\wh{\boldsymbol{f}}_{\calS_1}^{(1)}-\boldsymbol{f}_{0\calS_1}^{(1)}) \right\|_{\infty} > \bar{G} \right) \\
    & \le \pr\left( \left\| \boldsymbol{r}_{\calS_1} \right\|_{\infty} > \bar\lambda_2\right) + \pr\left(\frac{\sqrt{q_1}}{ \bar\lambda_2} \vertiii{\All - \Anll}_{2,2}>1\right).
\end{split}
\label{equ:tail_bound_final_2}
\end{align}

We show that with $\bar\lambda_2 >  D_5^* (\sigma+1) \tau^{1/2}  \left( n^{-1} \log q_1 \right)^{1/2}$, we have
\begin{align}
\mathbb{P} \left(   \left\| \boldsymbol{r}_{\calS_1}  \right\|_{\infty} \ge \bar\lambda_2 \right)
\le   \exp \left( - D^{(3)} \bar\lambda_2^2 n  \right) 
\label{equ:tail_bound_3}
\end{align}
holds for some $D^{(3)} < D_6^* \left( (\sigma + 1)^{2} \tau \right)^{-1} $ where $D_5^*$ and $D_6^*$ are universal constants.
Meanwhile, 
suppose $\bar\rho_1$ is the largest eigenvalue of $\All$, then
\begin{align}
\pr \left( \sqrt{q_1} \vertiii{\All - \Anll}_{2,2} > u \right) \le \exp\left\{ -\frac{u^2 n}{C^2 \bar\rho_1^2 q_1}  \right\} \label{equ:tail_bound_4}
\end{align}
holds for some constant $C>0$, as long as $C$ and $q_1$ satisfy
\begin{align*} 
\frac{u^2}{C^2 \bar\rho_1^2} < q_1 \le  \sqrt{\frac{u^2 n }{\tau C^2 \bar\rho_1}}.
\end{align*}
The proof of \eqref{equ:tail_bound_3} and \eqref{equ:tail_bound_4} can be found in Lemmas 5 and 6 of \cite{guo2024rkhs}.

Finally, bound the rhs of \eqref{equ:tail_bound_final_2} using Lemmas \eqref{equ:tail_bound_3} and \eqref{equ:tail_bound_4} and note that it is dominated by the expression in \eqref{equ:vs_consistency_rate1}
under Condition \eqref{equ:bar_lambda_12_bound_1}.

\paragraph{B.3. Proof of the Remark 4(iii)}

Remark~4(iii) considers an ultra-high-dimensional FLM regime in which $\log p = O\bigl(n^{1-2\varsigma}\bigr)$ and $q \asymp n^{\varsigma}$ with $0 < \varsigma < \tfrac14.$
We provide additional details on the minimum rate requirements for the tuning parameters \(\lambda_1\), \(\lambda_2\), \(\bar\lambda_1\), and \(\bar\lambda_2\), as well as for the minimum detectable signal levels \(G\) and \(\bar G\).

\noindent\textbf{The rate for $\lambda_1$ and $\lambda_2$:}
By \eqref{equ:bar_lambda_12_bound_1} in Theorem~\ref{thm:2},
\[
\lambda_1 \gtrsim \left(\frac{\log(p-q)}{n}\right)^{1/2}
\asymp n^{-\varsigma}.
\]
Similarly,
\[
\lambda_2 \gtrsim \max\left\{\frac{q\log(p-q)}{n},\left(\frac{q^2}{n}\right)^{1/2}\right\}
\asymp \max\left\{n^{-\varsigma},\, n^{\varsigma-\frac12}\right\}.
\]
Since \(\varsigma<1/4\), we have \(n^{-\varsigma}\ge n^{\varsigma-1/2}\), and hence $\lambda_2 \gtrsim n^{-\varsigma}.$

\noindent\textbf{The rates for \(\bar\lambda_1\) and \(\bar\lambda_2\):}
By \eqref{equ:bar_lambda_12_bound_1}, it suffices to choose
\[
\bar\lambda_2 \gtrsim \max\left\{\frac{q\log q}{n},\left(\frac{q^2}{n}\right)^{1/2}\right\},
\]
Since \(q\asymp n^\varsigma\), we have
\[
\bar\lambda_2
\gtrsim
\max\left\{
O\!\left(n^{\varsigma-1}\log n\right),\,
O\!\left(n^{\varsigma-\frac12}\right)
\right\}
\asymp n^{\varsigma-\frac12}.
\]

Similarly,
\[
\bar\lambda_1 \gtrsim \left(\frac{\log q}{n}\right)^{1/2}
\asymp n^{-1/2}(\log n)^{1/2}.
\]
Because \(\bar\lambda_1/\bar\lambda_2 \ge c\) for some constant \(c>0\), \(\bar\lambda_1\) cannot be of smaller order than \(\bar\lambda_2\). Therefore,
\[
\bar\lambda_1 \gtrsim \max\left\{n^{-1/2}(\log n)^{1/2},\, n^{\varsigma-\frac12}\right\}
= n^{\varsigma-\frac12}.
\]

\noindent\textbf{The rates for \(G\) and \(\bar G\):}
By Conditions~\ref{ass:a5} and~\ref{ass:a9},
\[
G \asymp \lambda_1\lambda_2^{-1/2}+\lambda_2^{1/2},
\qquad
\bar G \asymp \bar\lambda_1\bar\lambda_2^{-1/2}+\bar\lambda_2^{1/2}.
\]
Substituting \(\lambda_1 \asymp \lambda_2 \asymp n^{-\varsigma}\) and
\(\bar\lambda_1 \asymp \bar\lambda_2 \asymp n^{\varsigma-\frac12}\), we obtain
\[
G \asymp n^{- \frac{\varsigma}{2} },
\qquad
\bar G \asymp n^{\frac{\varsigma}{2}-\frac14}.
\]

\subsection*{C. Additional Details for the Data Application}

\paragraph{C.1. Accounting for dependence in the response}

\begin{figure}[htbp]
    \centering
    \begin{subfigure}[b]{0.488\textwidth}
        \centering
        \includegraphics[width=\textwidth]{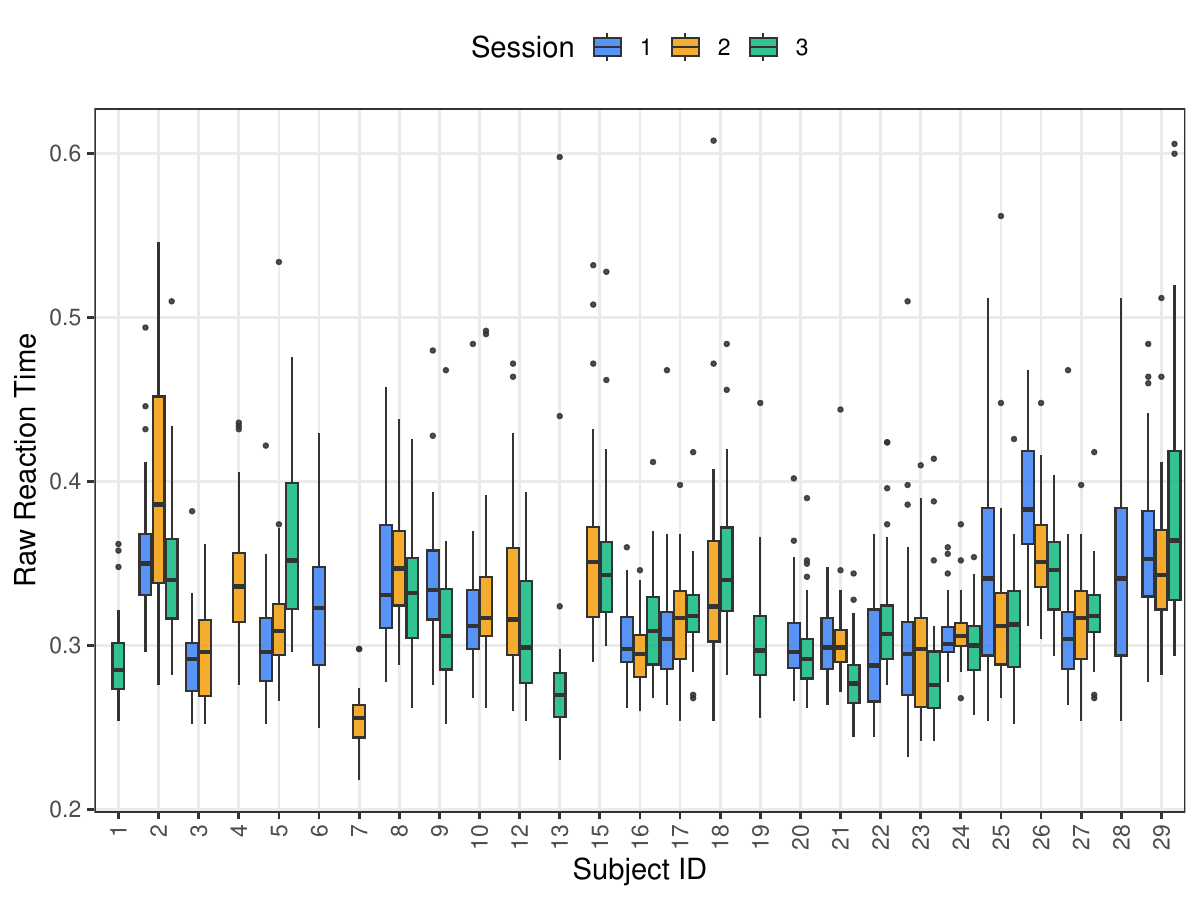}
        \caption{Raw RT}
    \end{subfigure}
    \hfill
    \begin{subfigure}[b]{0.488\textwidth}
        \centering
        \includegraphics[width=\textwidth]{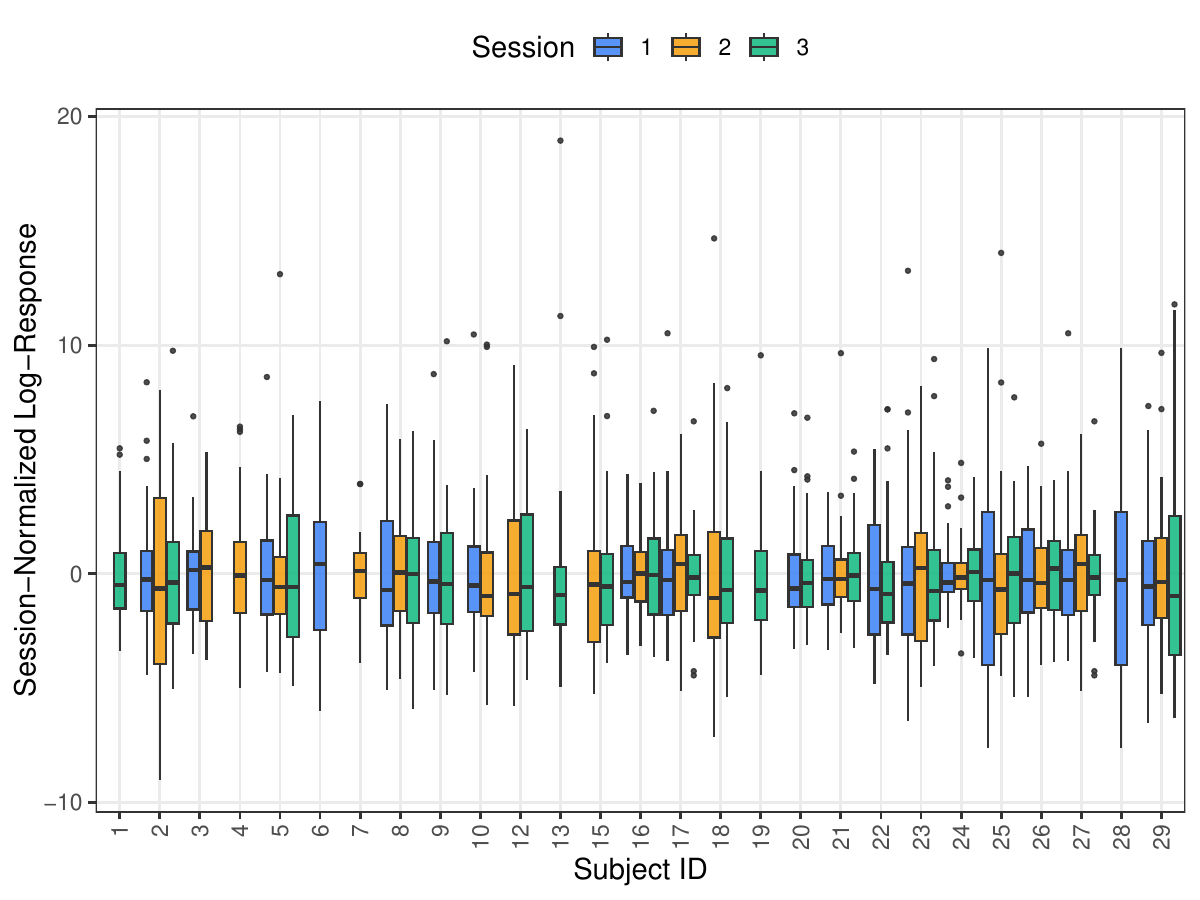}
        \caption{Session-normalized log-RT}
    \end{subfigure}
    \caption{Response time (RT) distributions by subject and session.}
    \label{fig:rt_dist}
\end{figure}

In the application, the observations are not independently collected. For example, the response time (RT) recorded for a subject at a given trial within a session may reflect subject-specific and session-specific effects, as well as temporal dependence across trials within the same session.
To illustrate this, Figure~\ref{fig:rt_dist}(a) displays boxplots of the raw RT values grouped by subject and session.

To reduce subject- and session-specific mean effects, we consider session-normalized log response RTs. Specifically, we first take the logarithm of the RTs and then center them within each session.
We also center the functional predictors within each session across trials. After this normalization, the subject- and session-level effects are substantially reduced, as illustrated in Figure~\ref{fig:rt_dist}(b).

\begin{figure}[htbp]
    \centering
    \begin{subfigure}[b]{0.488\textwidth}
        \centering
        \includegraphics[width=\textwidth]{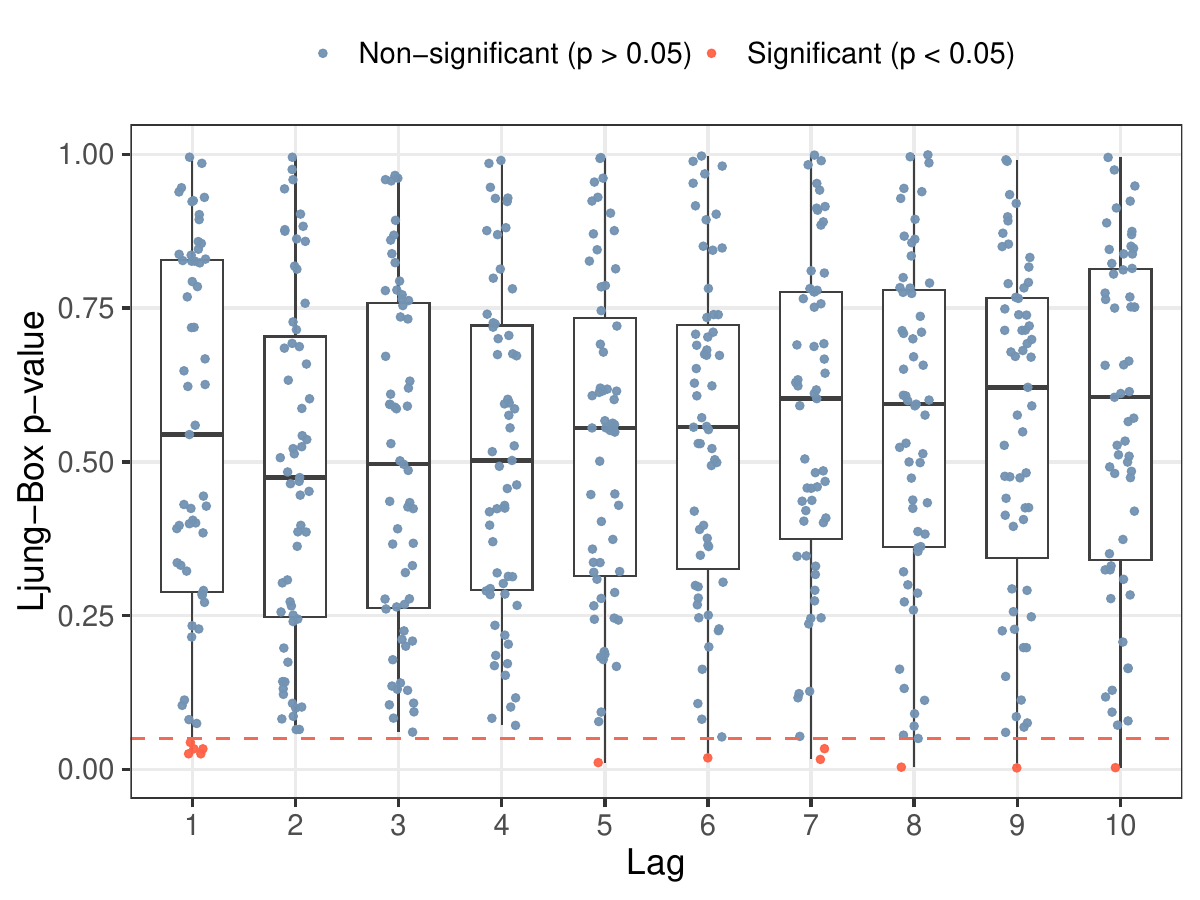}
        \caption{Distribution of Ljung--Box test p-values.}
    \end{subfigure}
    \hfill
    \begin{subfigure}[b]{0.488\textwidth}
        \centering
        \includegraphics[width=\textwidth]{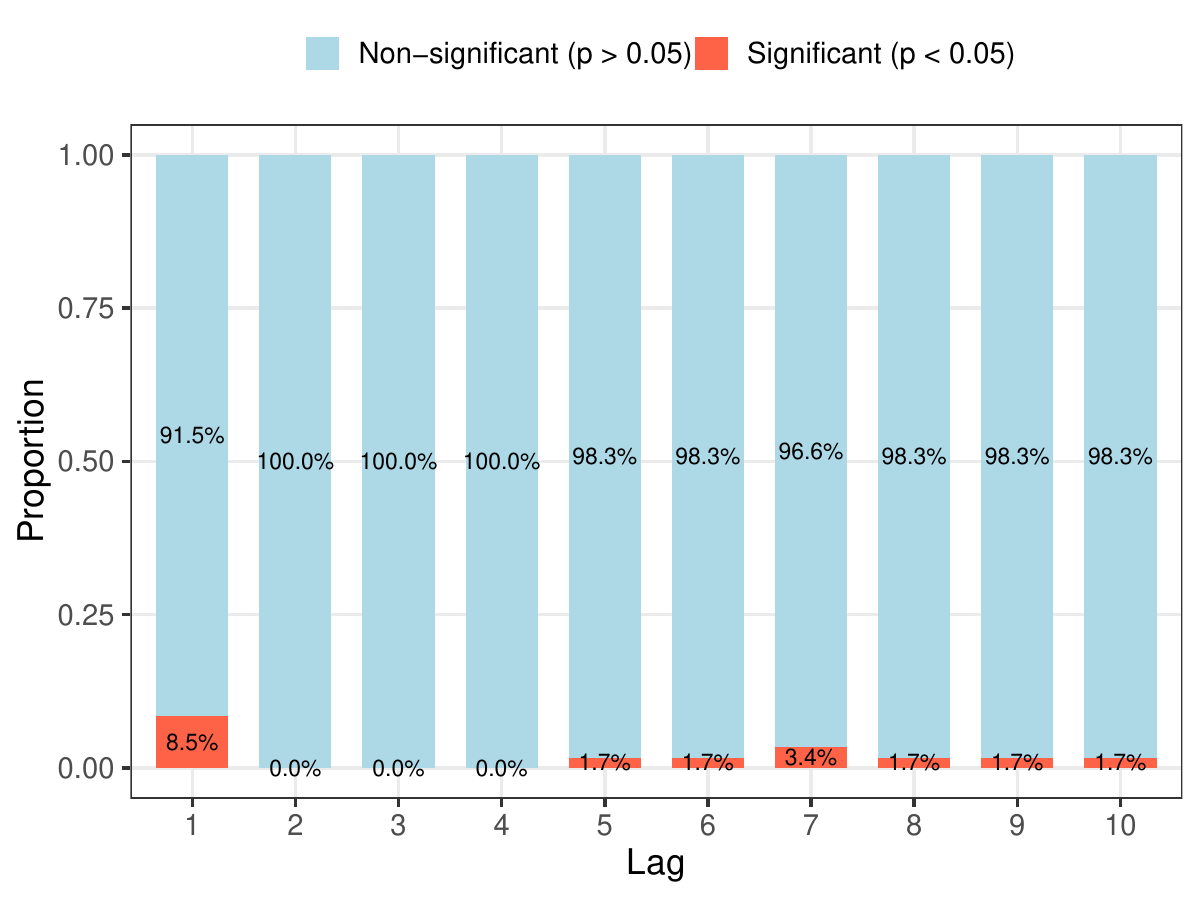}
        \caption{Proportion of significant Ljung--Box test results.}
    \end{subfigure}
    \caption{Residual dependence diagnostics based on the Ljung--Box test. In panel (a), each dot represents the p-value of the test for a single session, and red dots denote sessions with p-values less than 0.05.}
    \label{fig:res_dep}
\end{figure}

We conduct a residual analysis to assess the remaining within-session dependence after preprocessing and model fitting.
First, for each session, we apply the Ljung--Box test \citep{ljung1978measure} at lags one through ten. The resulting p-values are displayed in Figure~\ref{fig:res_dep}(a), and Figure~\ref{fig:res_dep}(b) summarizes, for each lag, the proportion of sessions with significant p-values. Overall, for lags one through ten, most sessions do not show significant results, suggesting that the residual temporal dependence is weak.

Second, we fit a linear mixed-effects model to the residuals to assess any remaining subject-level and session-within-subject variation. The corresponding tests provide no evidence of significant subject or subject-within-session random effects after the response transformation.

These analyses suggest that, after transforming the response, the proposed procedure remains reasonable for the present application.

\paragraph{C.2. Dependence across EEG channels}

\begin{figure}[ht]
    \centering
    \includegraphics[width=0.65\textwidth]{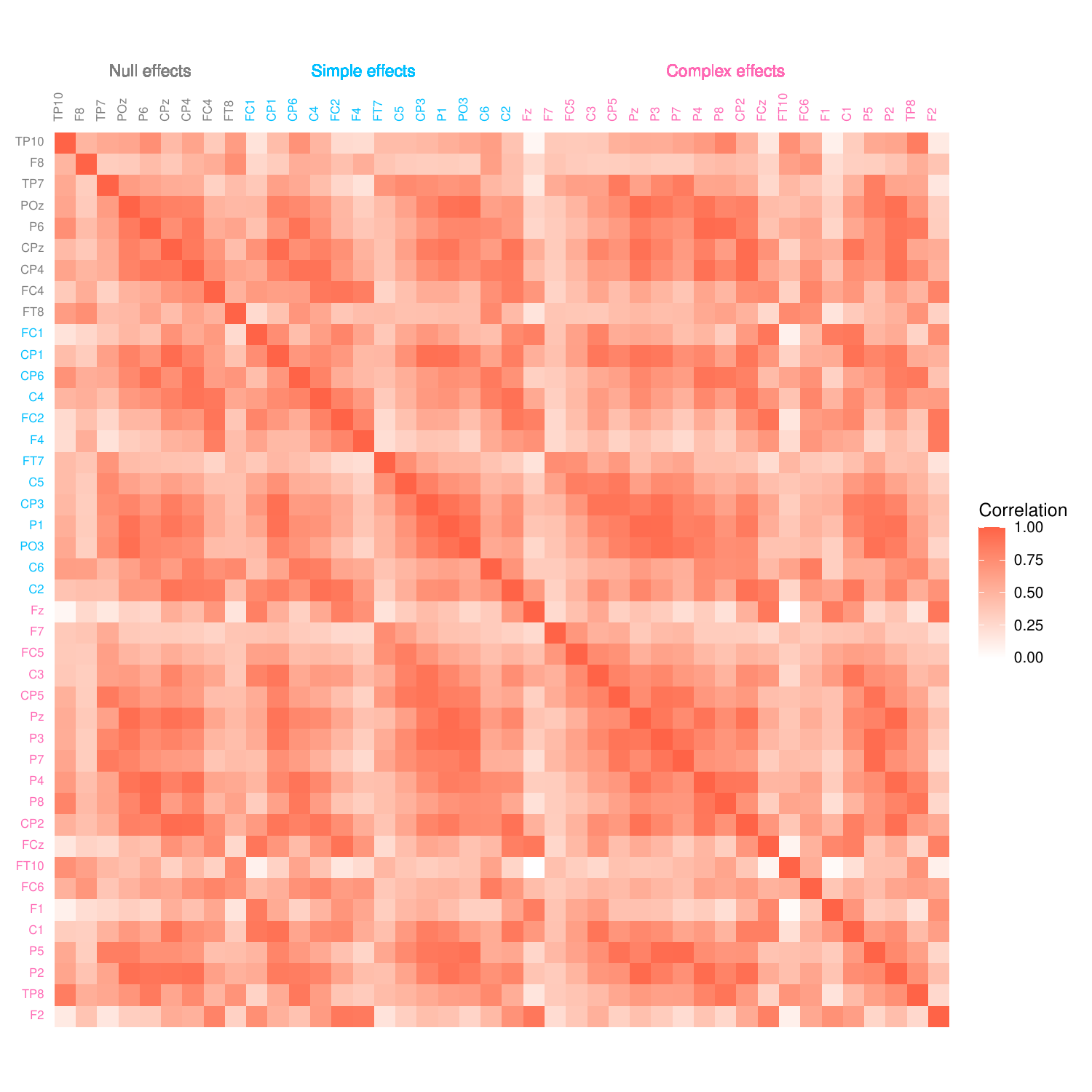}
    \caption{Average functional connectivity across all trials, measured by Pearson correlation between pairs of EEG channels over time. The heatmap indicates substantial dependence among EEG channels, especially for nearby locations. Channel labels are color coded by estimated effect type: null, simple, or complex.}
    \label{fig:connectivity}
\end{figure}

We compute the average functional connectivity across all trials, where functional connectivity is measured by the Pearson correlation between pairs of EEG channels over time. The resulting heatmap in Figure~\ref{fig:connectivity} shows substantial dependence among channels, particularly between spatially nearby locations.

Such strong cross-channel dependence makes this application considerably more challenging, which is consistent with the sufficient conditions required in our main theoretical results. In particular, strong dependence can reduce the separation between relevant and irrelevant channels, which may help explain why many channels are estimated to have nonzero effects in Figure~\ref{fig:application1}. It may also make the form identification between simple and complex effects more difficult.
For this reason, the results of this application should be interpreted with additional caution.

Nevertheless, the strong cross-channel dependence in this application does not necessarily invalidate the proposed procedure.
First, the correlation conditions in our theoretical analysis are sufficient rather than necessary, so their violation only indicates that the current theory does not fully cover this setting. Second, our penalty has an elastic-net-type form, which is known to behave more favorably in the presence of highly correlated predictors due to its grouping effect. Third, in EEG applications, substantial dependence among nearby channels is scientifically expected, so the selection of multiple correlated channels may still reflect meaningful signal structure rather than a failure of the method. 

\paragraph{C.3. Raw EEG signals and basic preprocessing}

       \begin{figure}[ht]
        \centering
        \includegraphics[width=0.7\textwidth]{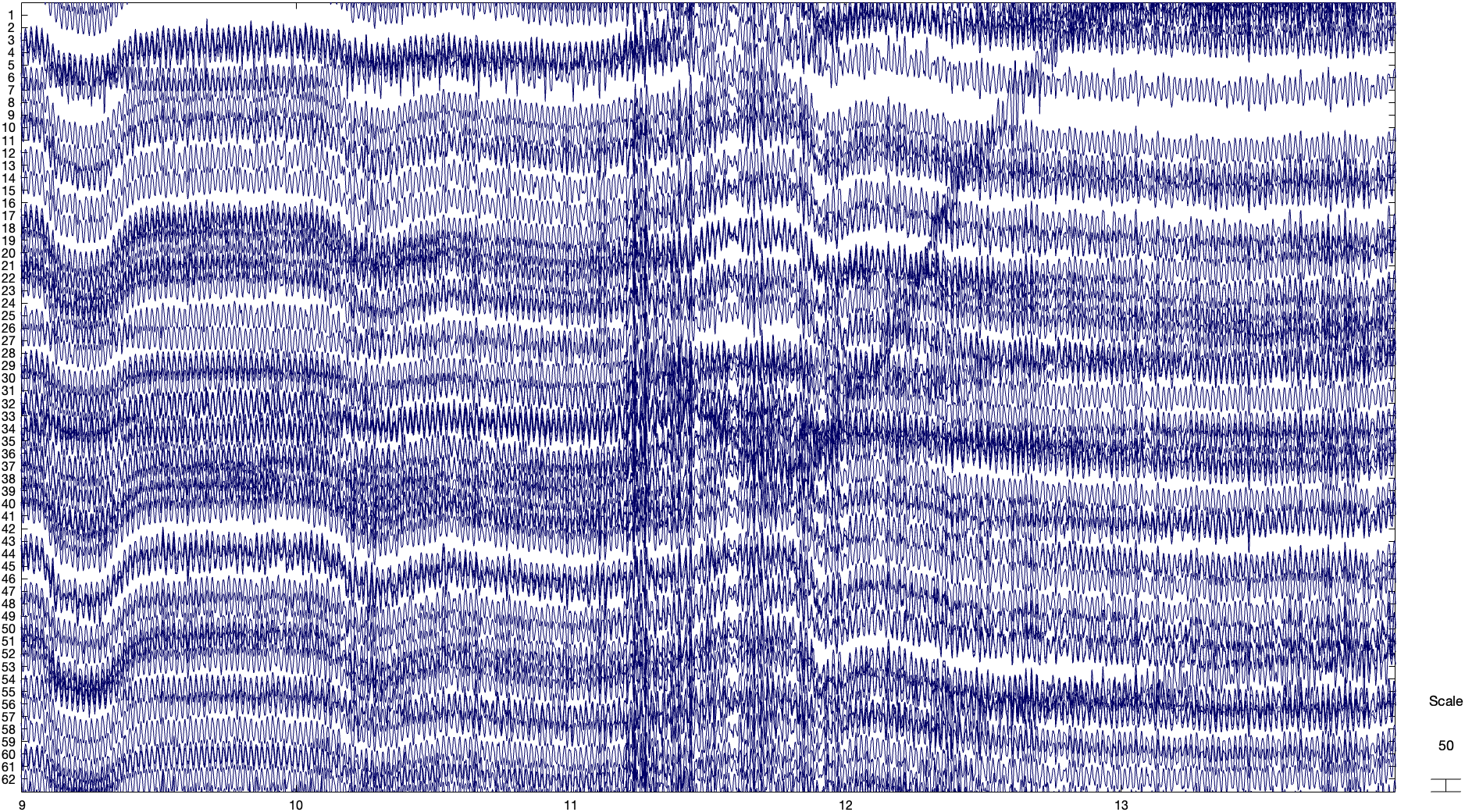}
    \caption{Visualization of the raw EEG signal and the main preprocessing steps.}
    \label{fig:eeg_example}
\end{figure}

Figure~\ref{fig:eeg_example} displays the raw EEG signals. The raw signals exhibit substantial high-frequency noise, low-frequency fluctuations, bad channels, and pronounced artifacts, likely arising from eye blinks or body movement. Such components are typically not of primary interest in standard EEG analysis and may substantially contaminate the signals. As a result, raw EEG data are generally not suitable for direct analysis and usually require at least basic preprocessing to improve reliability and reduce artifactual contamination \citep{delorme2004eeglab, bigdely2015prep}. In our analysis, we applied only essential preprocessing steps and avoided more aggressive procedures, so that the resulting EEG signals still retain substantial temporal complexity.

\vskip 0.2in
\bibliography{ref}

\end{document}